%% file: ms.tex
\documentclass[10pt, preprint]{aastex}
\usepackage{epsfig}
\usepackage{longtable}

\shorttitle{AGN Obscuration and Colors}
\shortauthors{Ammons et al.}

\begin{document}

\title{AGN Unification at $z\sim1$:  $u - R$ Colors and Gradients in X-ray AGN Hosts} 

\author{S. Mark Ammons\altaffilmark{1, \dag}, David J. V. Rosario\altaffilmark{2}, David C. Koo\altaffilmark{3}, Aaron A. Dutton\altaffilmark{4}, Jason Melbourne\altaffilmark{5}, Claire E. Max\altaffilmark{3}, Mark Mozena\altaffilmark{3}, Dale D. Kocevski\altaffilmark{3}, Elizabeth J. McGrath\altaffilmark{3}, Rychard J. Bouwens\altaffilmark{6}, and Daniel K. Magee\altaffilmark{3}}

\altaffiltext{1}{present address:  Steward Observatory, University of Arizona, 933 Cherry Ave., Tucson, AZ  85721, ammons@as.arizona.edu}
\altaffiltext{\dag}{Hubble Fellow}
\altaffiltext{2}{present address:  Max-Planck-Institut fŸr extraterrestrische Physik (MPE), Giessenbachstr.1, 85748 Garching, Germany, rosario@ucolick.org}
\altaffiltext{3}{present address:  University of California, Santa Cruz, 1156 High St., Santa Cruz, CA  95064, koo, max, mmozena, kocevski, mcgrath, magee@ucolick.org}
\altaffiltext{4}{present address:  Dept. of Physics and Astronomy, University of Victoria, Victoria, BC, V8P 5C2, Canada, dutton@uvic.ca}
\altaffiltext{5}{present address:  California Institute of Technology, MS 301-17, Pasadena, CA, 91125, jmel@caltech.edu}
\altaffiltext{6}{present address:  Leiden Observatory, Leiden University, NL-2300 RA Leiden, The Netherlands, bouwens@ucolick.org}

\begin{abstract}
We present uncontaminated rest-frame $u-R$ colors of 78 X-ray-selected AGN hosts at $0.5 < z < 1.5$ in the Chandra Deep Fields measured with HST/ACS/NICMOS and VLT/ISAAC imaging.  We also present spatially-resolved $NUV-R$ color gradients for a subsample of AGN hosts imaged by HST/WFC3.  Integrated, uncorrected photometry is not reliable for comparing the mean properties of soft and hard AGN host galaxies at $z \sim 1$ due to color contamination from point-source AGN emission.  We use a cloning simulation to develop a calibration between concentration and this color contamination and use this to correct host galaxy colors.  

The mean $u - R$ color of the unobscured/soft hosts beyond $\sim6$ kpc is statistically equivalent to that of the obscured/hard hosts (the soft sources are $0.09 \pm 0.16$ magnitudes bluer).  Furthermore, the rest-frame $V-J$ colors of the obscured and unobscured hosts beyond $\sim6$ kpc are statistically equivalent, suggesting that the two populations have similar distributions of dust extinction.  For the WFC3/IR sample, the mean $NUV - R$ color gradients of unobscured and obscured sources differ by less than $\sim 0.5$ magnitudes for $r > 1.1$ kpc.  These three observations imply that AGN obscuration is uncorrelated with the star formation rate beyond $\sim1$ kpc.

These observations favor a unification scenario for intermediate-luminosity AGNs in which obscuration is determined geometrically. Scenarios in which the majority of intermediate-luminosity AGN at $z\sim1$ are undergoing rapid, galaxy-wide quenching due to AGN-driven feedback processes are disfavored. 
\end{abstract}

\section{INTRODUCTION}

The physical connection between embedded supermassive black holes and their host galaxies remains an active topic of research.  The discovery of correlations between bulge properties and black hole mass \citep[e.g., stellar mass and velocity dispersion,][]{mag98, fer00, geb00} imply that this connection spans an astonishing range of spatial scales;  the AU-sized accretion disk whose characteristics determine the black hole growth rate must receive information about the conditions of the galaxy on scales of kiloparsecs, or eight orders of magnitude larger.  These correlations between black hole mass and various bulge properties hint at simultaneous evolution of AGNs and their host galaxies.

The processes that are involved in the maintenance of the $M-\sigma$ relation may also govern the evolution of host galaxy morphology and color.  In the local universe, galaxies exhibit bimodal distributions of several critical parameters, including rest-frame optical color, star formation rate, and gas fraction \citep[][etc]{str01, kau03, bal04}.  Star formation must be suppressed significantly to explain the low star formation rates and the redness of early-type, passively evolving galaxies today.  There is circumstantial evidence for the participation of AGN in this suppression of star formation, or ``quenching,'' including their presence in the green valley of the galaxy color-magnitude diagram (CMD), intermediate between the blue cloud and red sequence \citep{kau03, nan07, sch09, car10}, although the AGN fraction does not appear to be higher in the green valley \citep{xue10}. 

Major mergers of gas-rich galaxies can result in the rapid growth of central black holes.  In this picture, the collision of gas clouds saps angular momentum, efficiently sending gas to the central regions of the galaxy \citep{her89}, where it is largely consumed in star formation and partially accreted onto one or more BHs.  In several models, the ``M-$\sigma$'' relation is maintained through self-regulating processes involved in galaxy and supermassive black hole (SMBH) growth \citep[][and references within]{hop05, dim05, hop09}.    This self-regulation may be modulated by star formation, in which momentum injection by Type Ia supernovae and radiation pressure from O stars disrupt further star formation \citep{mur05}.  However, the immense energies associated with gravitational accretion onto a SMBH dominate the energy budget of the galaxy; this suggests that the AGN itself may provide sufficient energy to remove gas and limit the growth of the BH \citep{sil98, hae98, kin03, dim05, hop05, spr05, hop06, men08}.   A further consequence of the most energetic AGN feedback processes may be the expulsion of a majority of the galactic gas \citep{san88, dim05, hop06}.  This quenching of star formation may be a mechanism by which AGN activity controls the evolution of galaxies from the blue cloud to the red sequence.  

One feature of a model that includes rapid, AGN-induced quenching is an evolution between obscured and unobscured AGN states \citep{hop06, men08}.  The ejection of gas/dust in the central core of the galaxy is quickly followed by gas removal on kpc-scales.  This process naturally produces a positive correlation between AGN obscuration and the mass fraction of young stars (a consequence of sufficient gas supply) in the host galaxies \citep{men08}.  Unobscured AGN are expected to reside in relatively gas-free host galaxies, cleared by a previous blowout phase.  These hosts are expected to be redder than the hosts of obscured AGN, which would have higher star-formation rates.  
	
It is not clear how this theoretical evolution between obscured and unobscured states is consistent with the observed AGN unification paradigm.  AGN unification schemes attempt to explain the properties of the many observed types of AGN with a common physical model.  The most successful unification picture postulates that the differences between Seyfert types in the local universe are due to variation of the orientation of the observer relative to a centralized dusty torus \citep{ant93, urr95}.  Since obscuration is solely determined by geometry in a unified scheme, there is no reason to expect differences between the properties of the host galaxies of obscured and unobscured AGN. 
	
Many comparisons of obscured and unobscured AGNs have been performed at low and high redshift, using both optical- and X-ray-based diagnostics of AGN obscuration and various measures of star formation rate or gas content.  Locally, obscured Seyferts are associated with significant starforming activity, in some cases more than comparable unobscured hosts \citep{cid95, gon01}.  \citet{lac07} finds that obscured Type II QSOs at redshifts of $0.3 < z < 0.8$ are associated with significant star forming activity.  However, using HI measurements, unobscured broad-line AGN hosts at low-redshift are observed to be gas-rich \citep{ho08}.  \citet{sch09} find no dependence of host galaxy color on obscuring column for a small sample of SWIFT BAT AGN \citep{geh04, tue08} in the local universe.  \citet{pag04} find stronger submillimeter emission in an X-ray obscured sample compared to an X-ray unobscured sample, implying a higher star formation rate in the obscured sample.  \citet{gil09} find that the clustering lengths of $z \sim 1$ X-ray selected AGNs with broad-lines and those that lack broad lines are similar.  They also note that the clustering length of the X-ray hard sources is statistically similar to that of the soft sources, although the errors are dominated by number statistics.  \citet{sha10} measure the star formation rates of X-ray selected AGNs in GOODS-North with Herschel far-infrared fluxes, finding no correlation with X-ray absorbing column.  
	
\citet{pie10b}, CP10 hereafter, uses optical colors of X-ray selected AGNs at $z \sim 0.7$ in AEGIS to compare hard and soft sources.  Using colors measured with both integrated and extended apertures, they find that unobscured sources are bluer than obscured sources in rest-frame $U-V.$  
 	
\subsection{Comparing the Colors and Color Gradients of Obscured and Unobscured AGN}
	
Generally, understanding the relationship between AGN host galaxy properties on kpc-scales and the properties of the AGN on smaller scales can be helpful in constraining the role of AGN in the evolution of galaxies.  We focus on comparing the host properties of obscured and unobscured sources, given the hypothesis that unobscured sources will reside in redder hosts in a feedback scenario involving AGN-driven blowouts.  We adopt a two-pronged approach:  We utilize two imaging datasets of different spatial resolution, allowing us to probe host properties on two spatial scales.  We first measure the rest-frame $u-R$ colors of a large sample of AGN host galaxies at high galactic radii (r $\sim6-12$ kpc) and search for correlations of color with AGN obscuration or hardness ratio.  The combination of optical and infrared colors is a powerful diagnostic of the presence of young stars \citep{gil02}, particularly rest-frame $u-R$, or observed $V-J$ at $z\sim0.8$.  This sample is limited in spatial resolution by the infrared imaging data ($0\farcs3$ - $0\farcs6$ for ISAAC imaging and $\sim 0\farcs25$ for NIC3).

We then turn to a smaller sample of AGN host galaxies imaged by the Wide Field Camera 3 Infrared Channel (WFC3/IR) on HST as part of the Early Release Science (ERS) demonstration program \citep{win10}.  The improved spatial resolution of $\sim0\farcs2$ enabled by WFC3 allows us to probe rest-frame $NUV-R$ color closer to the central regions of host galaxies at $r \sim 1$ kpc, where AGN-induced winds may be the most energetic and color effects potentially more drastic.

Section 2 identifies the VLT/ISAAC and ACS data we use in the Great Observatories Origins Deep Survey (GOODS) field South \citep{gia04} and the sample selection.  Section 3 details the methodology used to obtain uncontaminated aperture photometry of AGN host galaxies in these fields.  Section 4 describes a suite of cloning simulations used to check the analysis and derive error bars.  Section 5 presents results for all samples and Section 6 concludes.  We assume a flat cosmology throughout, with $H_0 = 70$ km/s/Mpc, $\Omega_M = 0.3$, and $\Omega_{\Lambda} = 0.7$.

\section{DATA}
This paper focuses on a set of $\sim78$ X-ray selected sources in the Great Observatories Origins Deep Survey (GOODS) field South \citep{gia04} and North \citep{ale03}, overlapping the Chandra Deep Fields (CDF).  The deep, 1 Megasecond Chandra exposure in the CDF-S and the 2 Ms exposure in the CDF-North have revealed hundreds of AGNs at $0.5 < z < 1.5$ \citep{gia02, bar03, szo04}.  Deep optical and near-infrared imaging from the HST Advanced Camera for Surveys (ACS) is available in four filters, F435W (B), F606W (V), F775W (\textit{i}), and F850LP (\textit{z}).  This paper makes use of deep J,H, and Ks imaging in the CDFS using the ISAAC instrument at the VLT.  The data reduction and depths are described in \citet{ret10b}.

We utilize deep $F110W$ and $F160W$ imaging publicly available in the GOODS-North and South fields.  These NIC-3 \citep{tho98} pointings have been acquired through a variety of programs, including Guest Observer, DDT and Parallel observing programs, totaling 1400 orbits of $F110W$ and $F160W$ imaging.  As part of an HST archival program, \citet{mag07} have reduced and mosaiced all NIC-3 imaging in GOODS.  This data reduction procedure is described fully in \citet{mag07}.  Further details regarding the use of this NIC-3 data set are given in section 4.2.1 of \citet{amm09}.  

The near-ultraviolet (NUV) filter is defined as used for the GALEX mission \citep{mart05, mor05, mor07}.   The $NUV-R$ color is defined as in \citet{sal07}.  All magnitudes are in AB units unless explicitly stated otherwise.  Selection criteria are enumerated in section \ref{sect:selection}.

\subsection{Near-Infrared ISAAC Imaging}
The ISAAC instrument on the VLT has been used to mosaic the ACS region of GOODS-South to median $5\sigma$ point source depths of $J_{AB}=25.2$, $H_{AB} = 24.7$, and $Ks_{AB} = 24.7$ \citep{ret10b}.  The spatial coverage in each of the bands is $172.4$, $159.6$, and $173.1$ arcmin$^2$ in J, H, and Ks respectively.  The exposure times varied from 3-8 hours in each band, totalling 1.3 Megaseconds.  This program was queue-scheduled to optimize image quality; the image FWHM's range from $0\farcs3$ to $0\farcs65.$   Fully reduced and photometrically calibrated data is available from the GOODS ISAAC webpage \citep{ret10a}.

\subsection{Near-Infrared NICMOS Imaging}
In this study, we use nearly 100 square arcminutes of NIC-3 $F110W$ and $F160W$ imaging in GOODS North and South.  These pointings, which have been acquired through a variety of observing programs and are publicly available, have been fully reduced and mosaiced by \citet{mag07}.  This reduction procedure includes basic calibration (removing the instrumental signature), cosmic-ray removal, treatment for post-SAA cosmic ray persistence and electronic ghosts, sky subtraction, non-linear count-rate correction, artifact masking, robust alignment and registration for large mosaics, weight map generation, and drizzling onto a final image frame \citep{mag07}.  Sensitivities vary across the GOODS fields due to differing levels of exposure.

\subsection{WFC3/IR ERS Imaging}	
We use public WFC3/IR imaging in the ERS region with the $F125W$ and $F160W$ filters.  The data were reduced using the standard calwf3 pipeline.  After pipeline calibration, a residual gain "pedestal" was noticed \citep[see][]{win10} and corrected through a multiplicative scaling of each quadrant.  Images were corrected for geometric distortion and drizzled onto a 90mas grid using the MultiDrizzle package \citep{koe02, koe07}.  Individual pointings were registered to a single astrometric grid using SWarp \citep{ber02}, which corrected for a small rotation offset in the astrometric solution from MultiDrizzle.  A final mosaic of the data was created with SWarp and a small residual sky gradient was removed.  The compact source sensitivity is 27.5 and 27.2 magnitudes AB for a $5\sigma$ detection in the $F125W$ and $F160W$ bands, respectively.

\subsection{Sample Selection}
\label{sect:selection}

The overall sample is composed of X-ray selected sources in the GOODS ACS regions with redshifts, as described below.  The overall sample is divided into subsamples for further analysis with cuts in the X-ray hardness ratio and X-ray luminosity.  The samples and selection criteria are enumerated in Table \ref{tab:tabulated_results}.

This sample is X-ray selected using the 1 Ms \textit{Chandra} Deep Field South (CDFS) exposure as described in \citet{gia02} and the 2 Ms CDF-N exposure \citep{ale03}.  All GOODS-South X-ray sources were chosen with a SExtractor \citep{ber96} S/N detection threshold of 2.1.  GOODS-North X-ray point sources are identified with WAVDETECT \citep{fre02} using a false-positive probability threshold of $10^{-7}$ in one of seven X-ray bands ranging from $0.5$ keV to $8.0$ keV.  This threshold is comparable to the GOODS-South S/N detection threshold of 2.1.  

Of 339 X-ray sources detected in GOODS-South \citep{gia02}, 191 unique sources are identified as extragalactic (non-stellar), lie in the ACS and ISAAC tiled region, and have photometric redshifts from COMBO-17 \citep{wol04, zhe04} or \citet{luo10} or spectroscopic redshifts from \citet{szo04} or \citet{sil10}.  Sources with extended X-ray emission are omitted from the sample.  A redshift cut isolating sources with $0.5 < z < 1.5$ and an optical magnitude cut of $R < 24$ trim the sample to 56 objects.   

Of the 503 X-ray point sources detected in GOODS-North \citep{ale03}, 284 have reliable spectroscopic redshifts in \citet{bar03}.  22 sources possess spectroscopic redshifts, fall in the redshift range $0.5 < z < 1.5,$ appear within the NIC-3 $F160W$ tiled region in GOODS-North, have $L_{X, 0.5-8 keV} > 10^{42}$ ergs s$^{-1}$, and have $R < 24.$  The median X-ray detection S/N between 0.5 and 8.0 keV for the remaining sample is 11.95 and the minimum is 4.91, above the detection threshold for the GOODS-South catalog.  Including AGN from both northern and southern fields, 78 AGN remain in the sample.  The identification numbers, redshifts, X-ray luminosities, hardness ratios, $u-R$ colors, and $U-B$ colors are given in Table \ref{tab:data}.  
 
92\% of the sample of 78 AGN hosts have secure spectroscopic redshifts.   We divide the sample at a hardness ratio of $-0.2$ to isolate obscured and unobscured sources.  Sources with $HR < -0.2$ are termed \textit{unobscured} or \textit{soft} and sources with $HR > -0.2$ are termed \textit{obscured} or \textit{hard.}  These labels are used to refer to these types of sources throughout this paper.  We use the hardness ratio as defined as $HR = (H-S) / (H+R)$ in \citet{gia02} and \citet{szo04}, where H and S are the net instrument count rates in the hard (2-10 keV) and the soft (0.5-2 keV) bands, respectively.  

Hardness ratio is expected to be an imperfect measure of AGN obscuration due to K-correction effects.  We assess this using a subsample of GOODS-South X-ray sources with full X-ray spectral fits available in \citet{toz06}.  A hardness ratio of $-0.2$ at $z \sim 1$ corresponds to a neutral hydrogen column density of $2 \times 10^{22}$ cm$^{-2}$.  The neutral hydrogen column corresponding to HR = $-0.2$ shifts by a factor of $\sim2.5$ from $z \sim 0.5$ to $z \sim 1.5$ due to the k-correction \citep{toz06}.  The range of hardness ratios that correspond to $2 \times 10^{22}$ cm$^{-2}$ over the redshift range $0.5 < z < 1.5$ is $-0.3 <$ HR $< -0.1$.  25\% of soft sources selected with HR $< -0.2$ have $N_H > 2 \times 10^{22}$ cm$^{-2}$; 18\% of hard sources selected with HR $> -0.2$ have $N_H < 2 \times 10^{22}$ cm$^{-2}$.  This low level of sample mixing does not affect the conclusions of this paper.

The final redshift ranges, mean redshifts, and numbers of sources in each subsample are indicated in Table \ref{tab:tabulated_results}.   Table \ref{tab:tabulated_results} lists three groups, each of which are further subdivided into hard and soft subsamples:  ``all'', ``lum'', and ``ERS.''   The ``all'' sample is selected according to the criteria listed above.  The ``lum'' group is a subsample of the ``all'' group and only includes objects more luminous than $L_{X, 0.5-10 keV} = 10^{43.5}$ ergs s$^{-1}.$  The ``ERS'' subsample only includes sources in the ERS region sampled by HST WFC3 imaging.   

\begin{figure}
  \caption{Redshift histograms for soft (blue line) and hard (red line) sources.  Soft and hard sources are divided at a hardness ratio of $-0.2.$  The mean redshifts of the two populations are shown as dashed lines.  The mean redshifts of the samples match when the redshift limits are trimmed as described in Section \ref{sect:selection}.  The redshift distributions of the soft and hard sources are broadly similar, both displaying peaks at known redshifts of $z \sim 0.7$ and $z \sim 1.0$.}
  \label{fig:redshift_histo}
  \begin{center}
    \includegraphics[width = 3.0in, height = 3.0in]{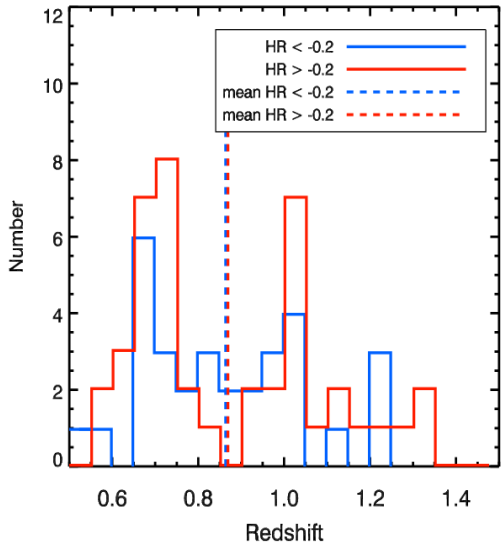}

 \end{center}
\end{figure}

To enable unbiased comparison of obscured and unobscured host properties, we cut the redshift ranges of the hard and soft subsamples to impose the mean redshifts of these two samples to match.  The redshift distribution of soft sources is generally weighted towards higher redshifts, so we trim the highest redshift sources from the soft subsamples.  Similarly, we trim low redshift sources from the hard samples.  As a result, both the mean redshifts and the mean X-ray luminosities of the soft and hard subsamples within each group match closely.  As shown in Figure \ref{fig:redshift_histo}, the trimmed redshift histograms of the two populations are broadly similar.  Note in Table \ref{tab:tabulated_results} that both the mean redshift and mean X-ray luminosity of the ``lum'' group is larger than the ``all'' group, as expected for a subsample of more luminous objects.  The mean redshift of the ``ERS'' group is significantly lower than that of the ``all'' group because the ERS region overlaps with a known galaxy overdensity at $z\sim0.73.$  The mean X-ray luminosity of the ``ERS-hard'' subsample and the ``ERS-soft'' subsample are significantly different, $10^{43.02}$ ergs s$^{-1}$ and $10^{43.54}$ ergs s$^{-1}$, respectively; this is due to two particularly X-ray luminous sources in the small ``ERS-soft'' subsample.  If these are removed, the mean colors of the ``ERS-soft'' sample do not change more than the $1\sigma$ measurement error.

\input{tab1.tex}

\input{tab2.tex}

\subsubsection{Color Magnitude Diagram}
\label{sect:CMD}

\begin{figure}
  \caption[Rest frame $u-R$ vs. $M_R$ color-magnitude diagram in AB magnitudes.]{Rest frame AB $u-R$ vs. $M_R$ color-magnitude diagram of the ``all'' sample, using integrated photometry.  Large circles denote strong ($L_{X, 0.5-10 keV} > 10^{43}$ ergs s$^{-1}$) sources and small circles denote weak ($L_{X, 0.5-10 keV} < 10^{43}$ ergs s$^{-1}$) sources.  Hard, obscured sources (HR $> -0.2$) are shown with overplotted squares and red colors.  Soft sources (HR $< -0.2$) are shown with blue colors.  Rest-frame values are estimated from observed photometry as in the text.  Both $u-R$ colors and $M_R$ are measured using a filled $3\farcs$ diameter circular aperture.  Colors and absolute magnitudes are uncorrected for central point source contamination.  Blue and red error bars denote the mean colors and luminosities for the soft and hard sources, respectively, with errors calculated as in Section \ref{sect:error}.  Small black circles denote the colors and magnitudes of the underlying galaxy population, selected as explained in the text.}
  \label{fig:UR_CMD}
  \begin{center}
    \includegraphics[width = 6.0in, height = 4.3in]{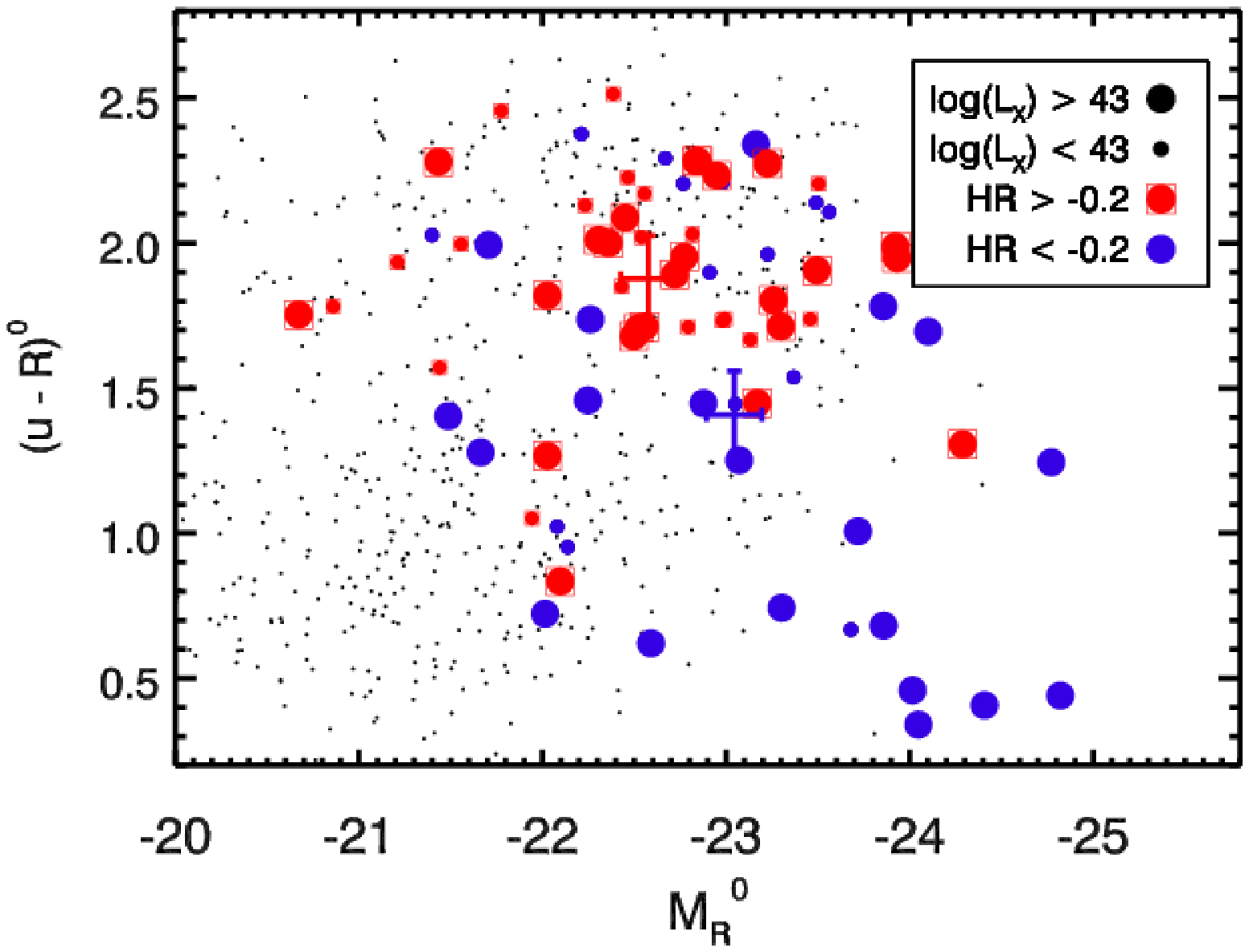}
 \end{center}
\end{figure}

Figure \ref{fig:UR_CMD} plots the rest-frame $u-R$ vs $M_R$ color magnitude diagrams (CMD) of the ``all'' sample.  The rest-frame photometry is obtained by performing $3\farcs$ filled-aperture photometry on all optical/NIR bands for all AGN, interpolating linearly through the SED, and sampling the de-redshifted SED at the appropriate filter wavelengths.  For comparison, rest-frame $u-R$ and $M_R$ measurements for the underlying galaxy population are plotted as small black circles for qualitative comparison.  This sample of $535$ control galaxies is selected from the GOODS MUSIC catalog \citep{san09} identically to the AGN sample (i.e., spectroscopic redshifts with $0.5 < z < 1.5$ and $R < 24$) except that X-ray sources with $L_X > 10^{42} $ ergs s$^{-1}$ are excluded.  Rest-frame colors are computed via interpolation through observed ACS/ISAAC \textit{bvizJHK} photometry, as described in Section \ref{sect:apphot}.  A $\sim0.15$ magnitude correction is added to the IR bands to correct for a difference in photometric aperture size between this sample and the MUSIC sample.  

In Figure \ref{fig:UR_CMD}, it is apparent that the AGN tend to reside in the most luminous systems.   Both red sequence galaxies ($u-R \sim 2.0$) and blue cloud galaxies ($u-R \sim 1.0$) are present in this sample \citep{sal07}.   The mean loci of the soft and hard sources are shown as blue and red error bars, respectively.  Using integrated photometry, it immediately appears that the soft sources are significantly bluer and more optically luminous than the hard sources.  However, integrated photometry is likely contaminated by the nonthermal emission produced by the AGN engine.  The next section introduces our method of measuring uncontaminated rest-frame host colors. 

\section{APERTURE PHOTOMETRY}
\label{sect:apphot}
We now measure aperture photometry with an elliptical annular aperture with inner de-projected radius of $0\farcs75$ and outer de-projected radius of $1\farcs5$, hereafter labeled the ``outer'' aperture.  We select a large inner radius of $0\farcs75$ that minimizes contamination from central point sources.  This dimension is oversized with respect to the spatial resolution of the near-IR imaging, which is set by ground-based seeing.  The outer radius of $1\farcs5$ is chosen to maximize the total signal-to-noise ratio over all bands for typical galaxy profiles of sersic index $n=2$.  

The axis ratios and position angles of the annuli are determined for each AGN host separately by manually fitting to isophotes in the ACS z-band image.  This is performed by lining up ellipses with major axis lengths of $3\farcs0$ with variable axis ratios and position angles until the nearest isophote is visually matched.  Only one annular aperture is used for each AGN host.

Rest-frame colors are computed from observed photometry with spline interpolations using IDL's \textit{interpol} function and known filter transmission functions.  The spline-fit SED is constrained to match the observed photometry upon integration with the appropriate filter curves.  Rest-frame magnitudes in $u$, $R$, and other bands are computed by integrating the de-redshifted SED with these filter functions.  We avoid using galaxy template spectra for fitting photometry, which can be problematic with the addition of non-thermal AGN continuum emission.  We test our technique by comparing rest-frame $u-R$ colors measured in a filled $2\farcs5$ diameter aperture with those computed by \citet{tay09} using SED-fitting techniques.  For $44$ isolated sources in GOODS-South that overlap the two samples, the $u-R$ colors match with a symmetric scatter of $\sigma = 0.15$ magnitudes and a minor systematic offset of $0.015$ magnitudes.  The scatter shrinks to $\sigma = 0.11$ magnitudes for sources not obviously dominated by central point sources, similar to the average level of measurement error in the sample.  We conclude that our technique is appropriate for measuring the average rest-frame colors of subsamples of galaxies.

The ``outer'' aperture is expected to be robust to contamination from central point sources associated with nonstellar AGN continua.  However, light from the central AGN may be scattered into the ``outer'' aperture by telescope diffraction, or the effect of a non-ideal Point Spread Function (PSF).  PSF contamination can be non-trivial for cases in which the AGN emission is dominant over the light from the host galaxy.  These cases require either a correction to the aperture photometry or special image processing to reveal the host galaxy.  We present two techniques for correcting aperture photometry for the effect of scattered AGN contamination.  Firstly, Section \ref{sect:apcorr} describes a method of estimating contamination from the PSF and subtracting its contribution from measured photometry.  Secondly, Section \ref{sect:CLEAN} describes the implementation of the CLEAN \citep{hog74, kee91} deconvolution algorithm to remove contamination directly from the images before measuring photometry.  Both methods requires excellent constraint of the PSF.

\subsection{Correcting ``Outer'' Aperture Colors for PSF Contamination}
\label{sect:apcorr}

Central embedded point sources contribute flux to annular photometric apertures via PSF scattering.   We approximate the contaminating flux $F_{contam}$ for a given source in a single band $S$ as 

$$F_{contam} \;=\; \frac{ \displaystyle\sum_{outer}\; F_{PSF, S}} {  \displaystyle\sum_{core} \;F_{PSF, S} } \times  \displaystyle\sum_{core}\; F_{image, S} $$

where $F_{image, S}$ is the flux distribution of the source in band $S$ and $F_{PSF, S}$ is the flux distribution of the centered PSF in band $S.$  $\Sigma_{outer}$ denotes a sum of fluxes over the ``outer'' annular aperture.  $\Sigma_{core}$ here denotes a sum over a filled, circular aperture with radius $0\farcs18$.  Note that the first fraction is similar to a concentration measurement, except that wider aperture is elliptical and annular.  The decontaminated flux is then given by 

$$F_{decontam} \;=\; \left( \displaystyle\sum_{outer}\; F_{image,S}\right) \; - F_{contam}$$

This method corrects for all PSF-scattered flux originating from within the central core aperture, no matter the source.  PSF-scattered flux originating from outside of the core aperture is not removed, but it is assumed that this light is intrinsically stellar.

\subsection{CLEAN Method}
\label{sect:CLEAN}
This section describes the use of the CLEAN algorithm to deconvolve galaxy images and allow estimation of intrinsic, uncontaminated photometry in a variety of photometric apertures. 

The CLEAN algorithm was originally designed to remove confusing sidelobe patterns in synthesized-beam radio interferometry at high resolution \citep{hog74} but has been modified to perform deconvolution of direct imaging \citep{kee91}.  CLEAN iteratively constructs a model of the instrinsic light distribution, returning both the model and the image residual.  In each iteration, CLEAN adds ``points'' (delta functions) to the model that correspond to the location of the maximum residual in the original image.  The intensity of the point is a fraction (gain) of the intensity of the maximum residual.  The convolution of the PSF with that point is subtracted from the original image.  The loop repeats until all of the flux from a specified region in the measured image has been removed or after a maximum number of iterations.  CLEAN preserves the photometric scale of the original image \citep{kee91}.

For this study, CLEAN is performed over a $4\farcs5$ box in both ACS and ISAAC images.  The chosen number of maximum iterations is 500 and the gain is $5\%.$  The algorithm is modified to find peaks using a cross-correlation of the PSF against the image rather than subtracting at the maximum pixel location.  	

The deconvolved image is the sum of the residual image and the CLEAN model.  Galaxy fluxes uncontaminated by the wings of any central point source can then be measured by performing aperture photometry directly on the deconvolved image with a specified aperture.

\subsection{Stacking Analyses}

Stacking sky-subtracted, reduced images of galaxies in various bands is helpful for revealing the mean fluxes and colors of various sub-populations, and serving as a check on other methods.   In this study, we stack images that have been masked to remove contaminating background sources and deconvolved with CLEAN.    We use observed $V$ and $J$ colors, which correspond to nearly $u$ and $R$ at $z\sim0.8$.  The gain in S/N realized by stacking is tempered by two major disadvantages:  (1) Observed-frame colors do not represent the mean rest-frame colors, as the sources lie at a variety of redshifts; and (2) Stacking requires the use of a single photometric aperture on a stacked image rather than a variety of apertures that correspond to the position angle and axis ratio of individual sources.  With these caveats in mind, we use a circular, annular photometric aperture of inner radius $0\farcs75$ and outer radius $1\farcs5$ to compute observed $V-J$ colors.   

\subsection{Sky Subtraction}

Meaurements of surface brightness at high galactic radii are highly sensitive to errors in sky subtraction.  The sky value for each band is measured by manually masking 3-sigma sources and computing the geometric mean of the values in a circular annulus with inner radius $4\farcs0$ and outer radius $6\farcs0.$  These radii were determined empirically by selecting empty regions in ACS and ISAAC fields and attempting to predict the sky value in an inner circle of $3\farcs0$ diameter; this combination of sky radii minimized the  errors in all ACS and ISAAC bands.  It is important to fix the annular radii across bands so as not to measure artificial colors from galactic contamination.   

Other methods of measuring sky were tested, including using sky measurements at varying radii to extrapolate the sky value inwards to zero radius.  The simple masked mean is superior to extrapolation.

\subsection{PSF Selection}
\label{sect:PSFs}
PSF reference stars are used to correct the annular photometry for the presence of bright, centralized point sources.  It is important that these PSF references be reliable to enable good recovery of underlying galaxy photometry. 

The ACS, NIC3, and WFC3 point spread functions are spatially and temporally stable compared to ground-based PSFs.  For each ACS, NIC3, or WFC3 band, images of bright unsaturated stars are taken from locations in the field.  The radial profiles of the stars are compared to TinyTim PSFs to verify that the sources are unresolved.  The PSFs are chosen to be in empty, isolated regions; any sources in the area are masked with replicated boxes of pixels taken from nearby empty regions with the area.  A sky subtraction is performed using the masked mean method described in the above section.  The PSFs are centered by sub-pixel interpolation.

For ISAAC PSF reference stars, the sources are verified as unresolved by comparing to ACS imaging.  In each ISAAC field, the brightest, isolated star with pixel values less than 10 times saturation level is selected.  Sky subtraction and centering are performed on these PSFs as described above.  The ACS and WFC3 PSF sizes are set to $3\farcs \times 3\farcs$, or the same size as the annulus used in aperture photometry, and the ISAAC and NIC3 PSFs are set to $7\farcs5 \times 7\farcs5$ to capture any light scattered to high radii.  

\subsection{Computing $NUV-R$ Color Gradients for ERS sample} 
\label{sect:color_grad_method}
The selection of the ERS sample of 34 X-ray sources is described in Section \ref{sect:selection}.  We select PSFs as described in Section \ref{sect:PSFs} and use the CLEAN algorithm to deconvolve the images as in Section  \ref{sect:CLEAN}.  To measure color gradients for individual sources, we first compute surface brightnesses in all observed bands with the CLEANed images for five concentric elliptical apertures of mean radii $0\farcs16$, $0\farcs33$, $0\farcs59$, $0\farcs88$, and $1\farcs27$.  The smallest of these apertures is filled and the larger four are annular.  The axis ratio and position angle of the ellipses is chosen as described in the first part of Section \ref{sect:apphot}.  For each aperture, we interpolate through observed $B$, $V$, $i$, $z$, $F125W$, and $F160W$ surface brightnesses to compute a rest-frame $NUV-R$ color.  For each source, we convert the apparent angles of arcseconds into physical units of kpc and interpolate the radius onto five points:  1.3, 2.67, 4.78, 7.13, and 10.3 kpc.  

\section{CHARACTERIZING MEASUREMENT ERROR WITH SIMULATIONS}

Multiple sources of error come into play in the measurements of the colors of AGN host galaxies at $z\sim1.$  First, host colors may be contaminated by the non-stellar colors of AGN emission.  In the case of integrated colors, photometric measurements with filled apertures directly sum AGN-related emission and stellar light.   In the case of ``outer'' colors measured with annular apertures, emission from the central AGN may still fall in the photometric aperture due to PSF scattering.  Second, our attempted method to correct for PSF contamination (described in \ref{sect:apcorr}) will be sensitive to PSF variation across the field.  Third, flux measurements in large annular apertures may be sensitive to sky subtraction errors, depending on the exact dimensions of the aperture.  

These potential errors motivate a detailed approach to estimating measurement errors.  Using a suite of cloning simulations, we characterize measurement error as function of the strength of a central point source and the intrinsic surface brightness of the galaxy host.  In these simulations, we modify images of local galaxies to approximate their appearances as if they were located at higher redshift.   One powerful advantage of cloning simulations is that we are able to measure the intrinsic surface brightness of an individual cloned galaxy with high S/N before PSF convolution.  As this information is difficult to obtain for high-redshift galaxies, cloning will allow us to measure errors at fainter host magnitudes than the alternative approach of adding point sources to neighboring galaxies at comparable redshifts to the AGN sample.  

\subsection{Implementing Cloning}

Sloan Digital Sky Survey \citep[SDSS,][]{yor00} images of ten local galaxies were downloaded from the SDSS website in \textit{u'g'r'i'z'} filters, mosaiced, and sky subtracted.  These ten galaxies were chosen because their distribution of inclination angle, half-light radius, and concentration are representative of the ``all'' sample.  The natural colors of these galaxies are not representative of the ``all'' sample, however, so the colors are manipulated to match the mean colors of the ``all'' sample during the cloning process.   The final reduced image sizes range from 27 kpc to 67 kpc, covering physical regions larger than the diameter of the most expansive photometric aperture at high redshift for this sample (25.4 kpc at $z = 1.5$).  The seeing in physical units ($\sim 20-50$ pc) is several orders of magnitude smaller than the effective image quality expected at high redshift ($\sim 500-4000$ pc).  Ultraviolet images were downloaded from the GALEX public website for these four galaxies in both near-UV (NUV) and far-UV (FUV) filters.  

Images of cloned galaxies are generated by inserting scaled copies of these SDSS and GALEX images into empty regions of ACS $B,V,i,z$, ISAAC $J,H,Ks$, and NIC3 $F110W$ and $F160W$ images.  The change in scales from the local universe to high redshift is calculated assuming a flat cosmology with $H_0 = 70$ km/s/Mpc, $\Omega_M = 0.3$, and $\Omega_{\Lambda} = 0.7$ with the IDL ``lumdist'' function.  For an individual cloned instance of a galaxy, the target redshift is generated randomly over the range $0.5 < z < 1.5$ and used to assign GALEX/SDSS filters to the ACS/ISAAC filters into which they redshift.  For a given redshift, the filter closest in wavelength is chosen.  At the lowest redshift of $z=0.5$, the SDSS $z'$ image is far bluer than the wavelength required to fill the ISAAC $Ks$ band image ($\sim1.5\; \mu $m).  However, the morphological K-correction between rest-frame $\sim0.9\; \mu m$ and $\sim1.5\; \mu m$  is expected to be small. 

Once source filters are assigned to target filters, the scaled copies are convolved with point spread functions drawn from the target images.  Each of 10 empty target locations in the ACS/ISAAC images is chosen nearby a reference star, which is used for convolution.  The reference star is not used for simulated observation in later steps.  The images and PSFs are supersampled by a factor of 3 for this convolution and brought back to the target pixel resolution.  The ISAAC point spread functions possess considerable noise, so 2-D Moffat fits of these PSFs are used for convolutions.  The radial profiles of the Moffat fits match the radial profiles of the fitted ISAAC PSFs to within 10\% at all radii less than $3\farcs0$ for all 10 target locations.  

For each cloned galaxy, the observed z-band surface brightness in the $0\farcs75 < r < 1\farcs5$ elliptical aperture is set randomly between $21$ and $31$ magnitudes per square arcsecond.   The amplitudes of the images in each of the other filters are chosen according to the mean rest-frame SED of the actual host galaxy population as measured from this data set ($(u-R)^0 \sim 1.6$; $(R-J)^0 \sim 1.1$).  The intrinsic host surface brightnesses (i.e., before convolution) in the elliptical aperture are computed and recorded for comparison later.  

To test the robustness of wide aperture photometry against contamination from AGN emission, a point source of random amplitude is added to the host galaxy center.  The {\it Point Source Fraction}, or ratio between the flux in the point source and the flux in the entire host galaxy, ranges from $0.1$ to $1000$.  

\subsection{Calculation of Critical Errors}
\label{sect:cal_errors}

The images of ten local galaxies are cloned 1600 times in each of 10 insertion locations in the ACS and ISAAC fields, 800 with no central point source, and 800 with central simulated point sources.  The redshifts and bolometric galaxy luminosity vary as described above.  For each clone, full aperture photometry is performed as described in Section \ref{sect:apcorr}, using an aperture correction to compute uncontaminated surface brightnesses.   We compare the measured surface brightness with the true, instrinsic surface brightness in the correct aperture before convolution with the PSF.  

\subsubsection{Galaxies without Central Point Sources}
\label{sect:sim_no_PS}

Here, we estimate photometric error in the ``outer'' aperture as a function of intrinsic host surface brightness using cloned galaxies that lack central point sources.   Given the measured surface brightnesses for the ``all'' sample, we estimate photometric errors and evaluate the reliability of the ``outer'' photometry for this sample.

\begin{figure}
  \caption{Plots of relative flux differences $(f_{meas} - f_{true}) / f_{meas}$ in ACS and ISAAC bands for cloned galaxies without central point sources, plotted as a function of true galaxy surface brightness in the ``outer'' aperture in units of AB magnitudes per square arcsecond.  Each filled circle denotes a separate cloned instance of a local galaxy.  Different colors identify the original local galaxy.  Gray dashed lines indicate the mean inferred systematic error as a function of surface brightness for these simulations.  Shaded regions denote the $1\sigma$ random error inferred from these simulations.  Red vertical lines plot the surface brightness at which the total inferred photometric error exceeds 0.4 magnitudes.  Solid black lines plot histograms of measured surface brightness for the ``all'' sample in each band.  The blue line in the upper left plot shows the histogram of measured $B$-band surface brightness for galaxies with redshifts $z < 0.638$, below which $B$-band measurements are required to compute a rest-frame $u$ flux.  Note that very few galaxies exceed $0.4$ magnitudes of error in any band, except for in the $B$-band.  However, for those galaxies that require an observed $B$ flux to compute a rest-frame $u$ flux, none exceed $0.4$ magnitudes of photometric error.}
  \label{fig:flux_dif}
  \begin{center}
    \includegraphics[width = 2.3in, height = 1.6in]{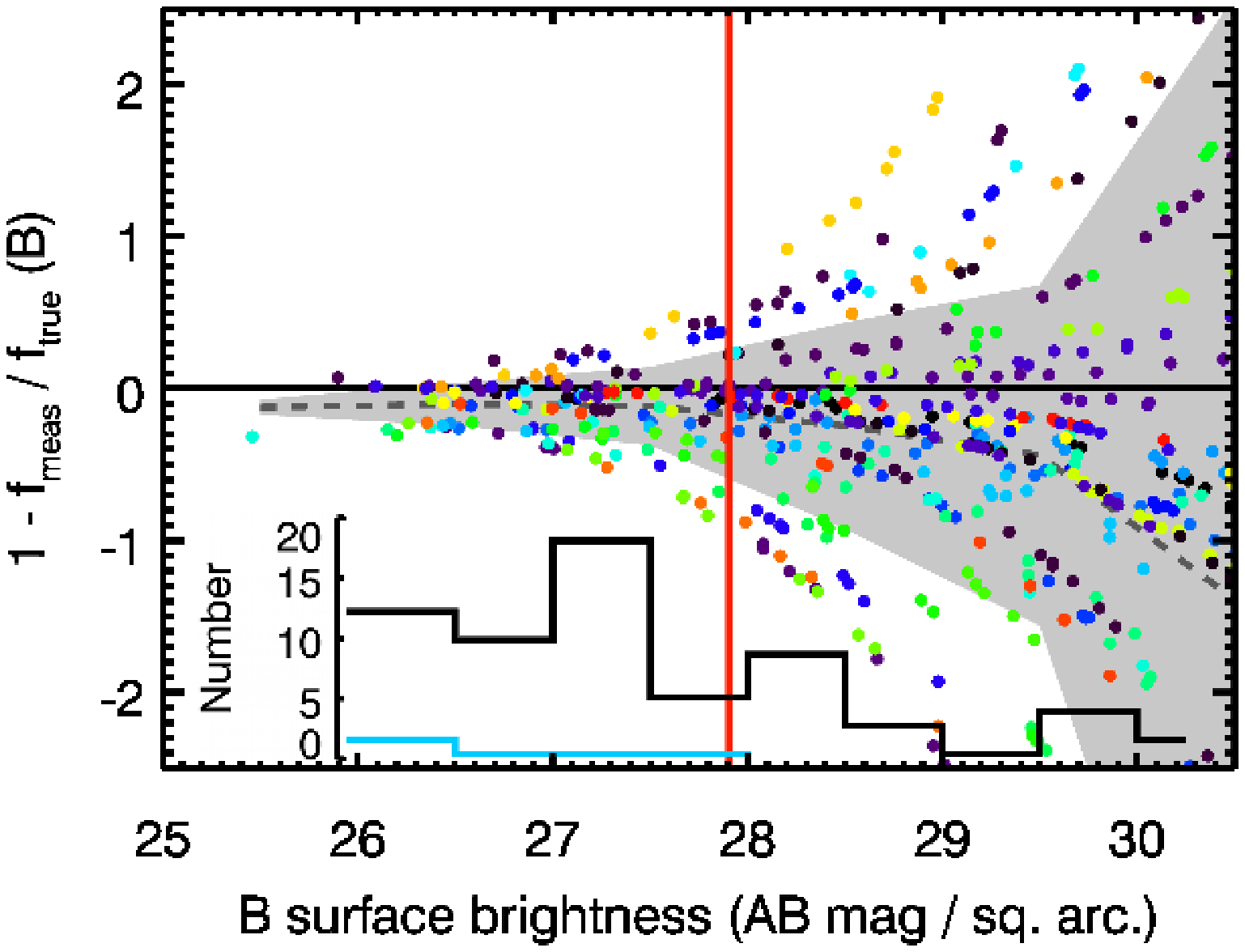}
    \includegraphics[width = 2.3in, height = 1.6in]{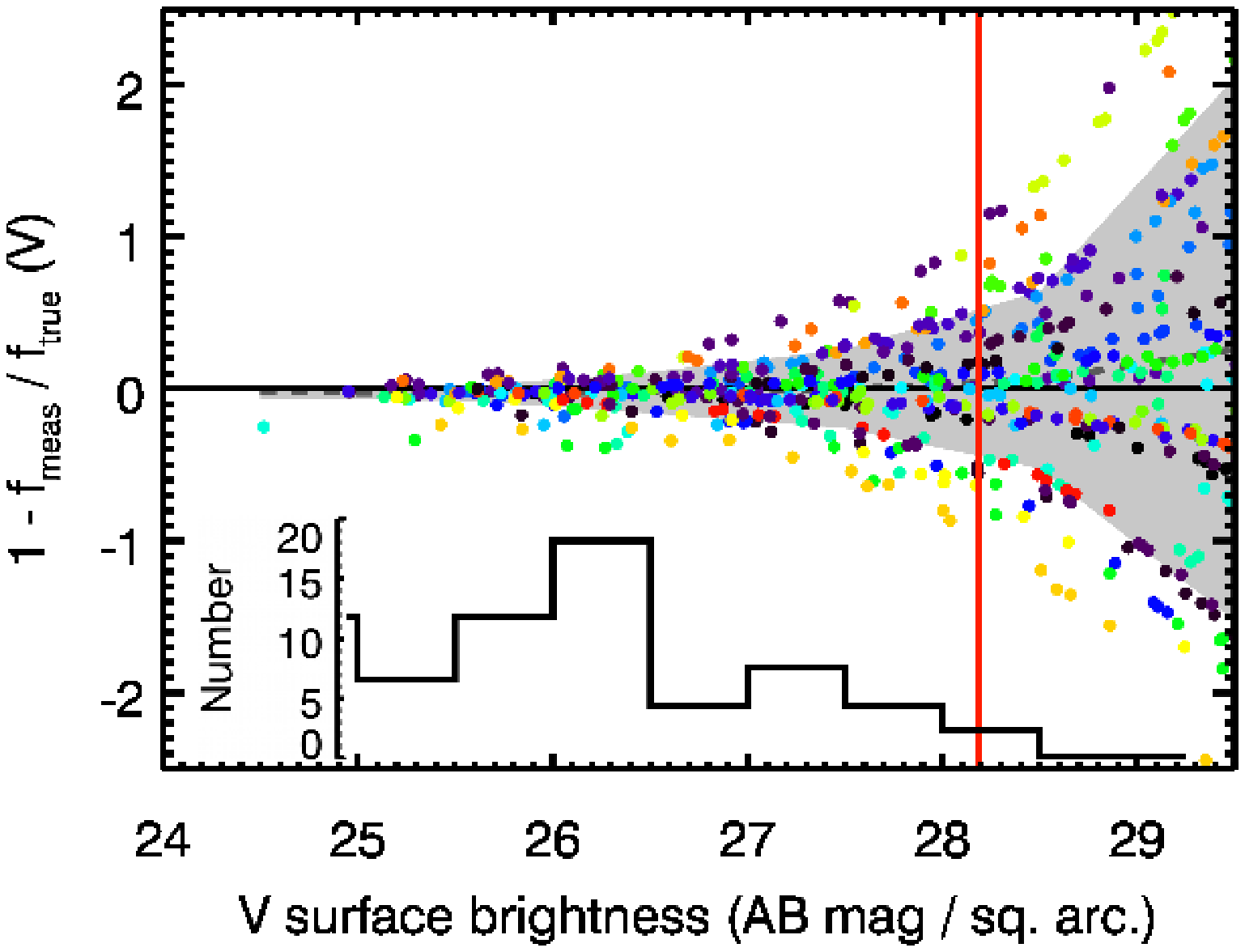}
    \includegraphics[width = 2.3in, height = 1.6in]{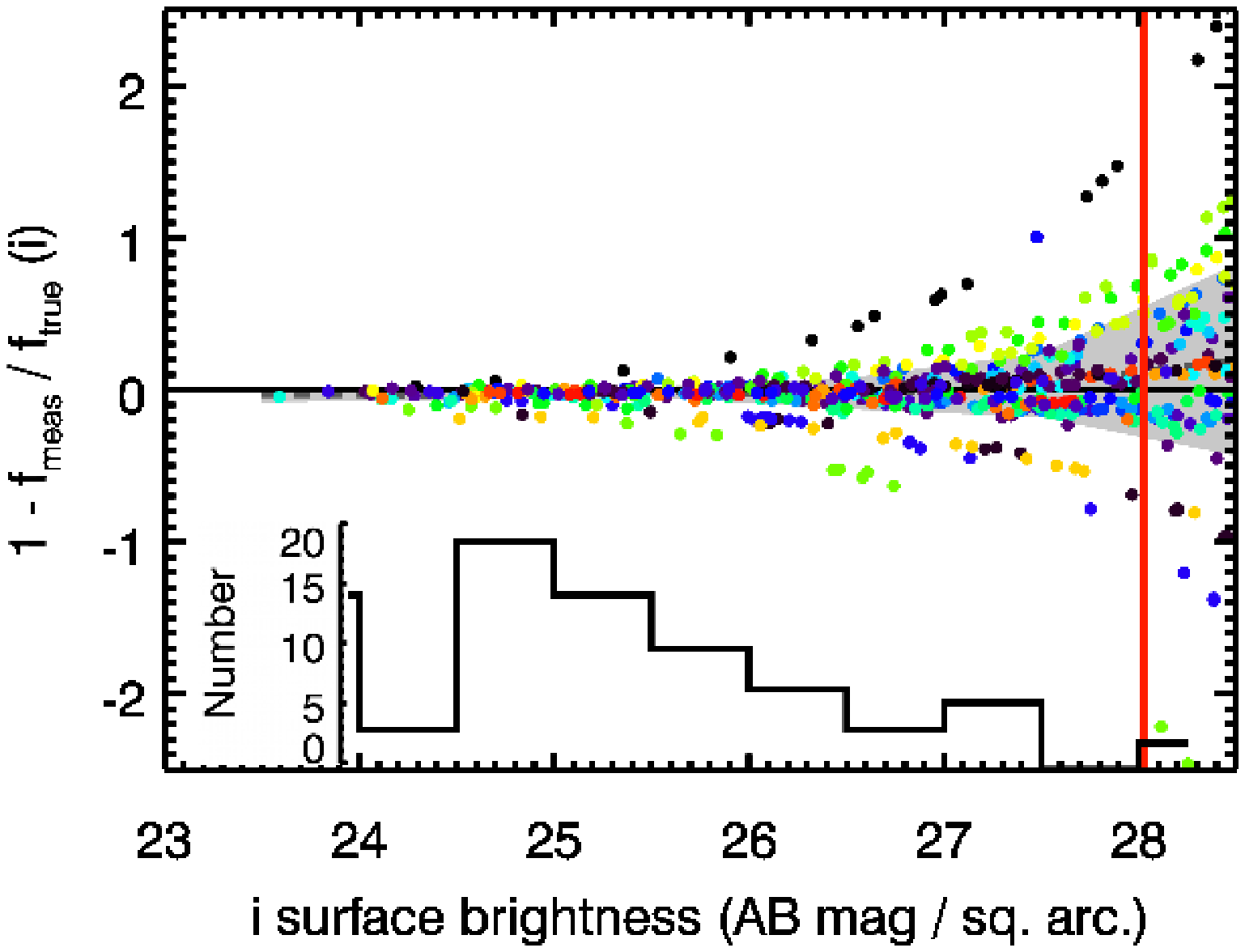}
    \includegraphics[width = 2.3in, height = 1.6in]{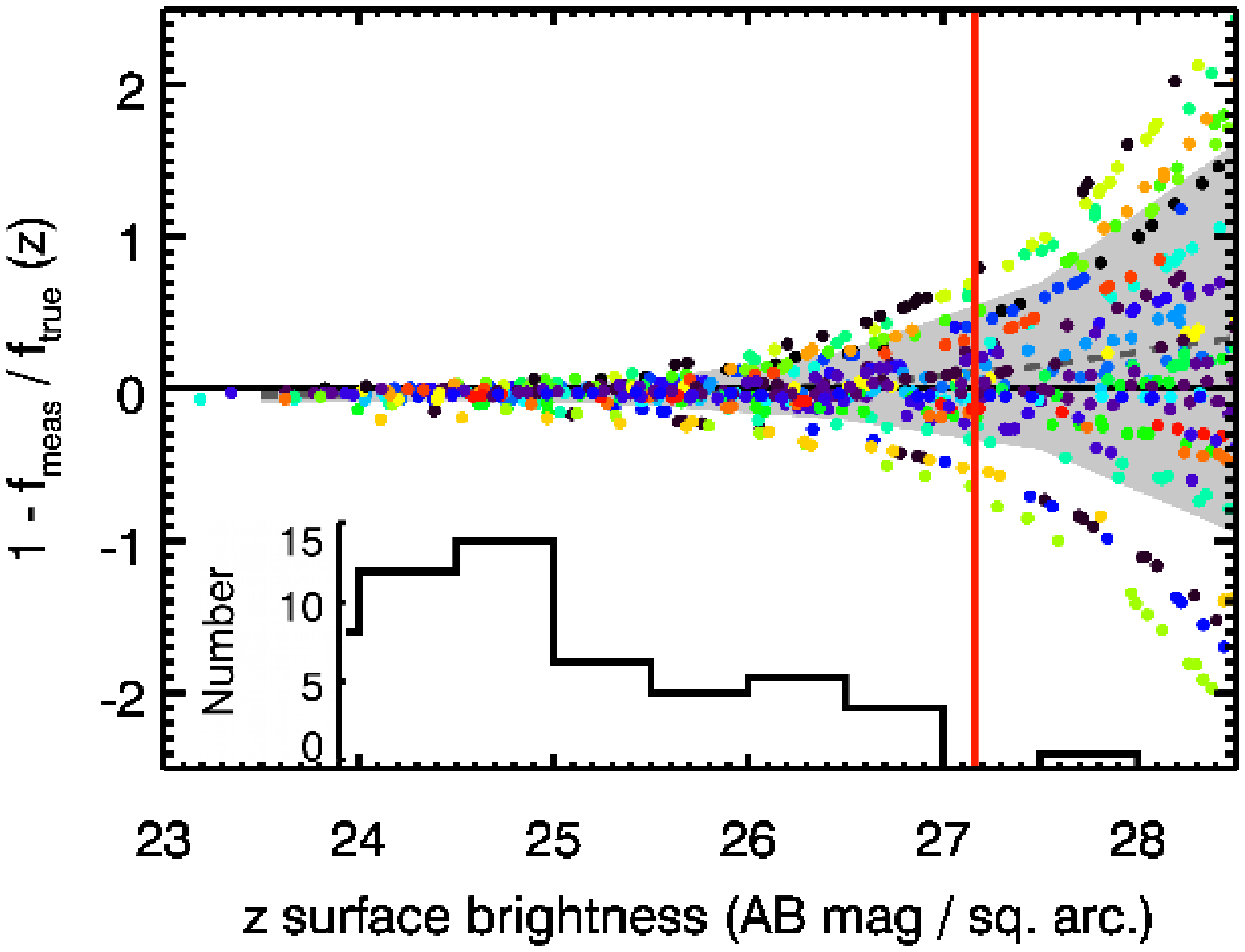}
    \includegraphics[width = 2.3in, height = 1.6in]{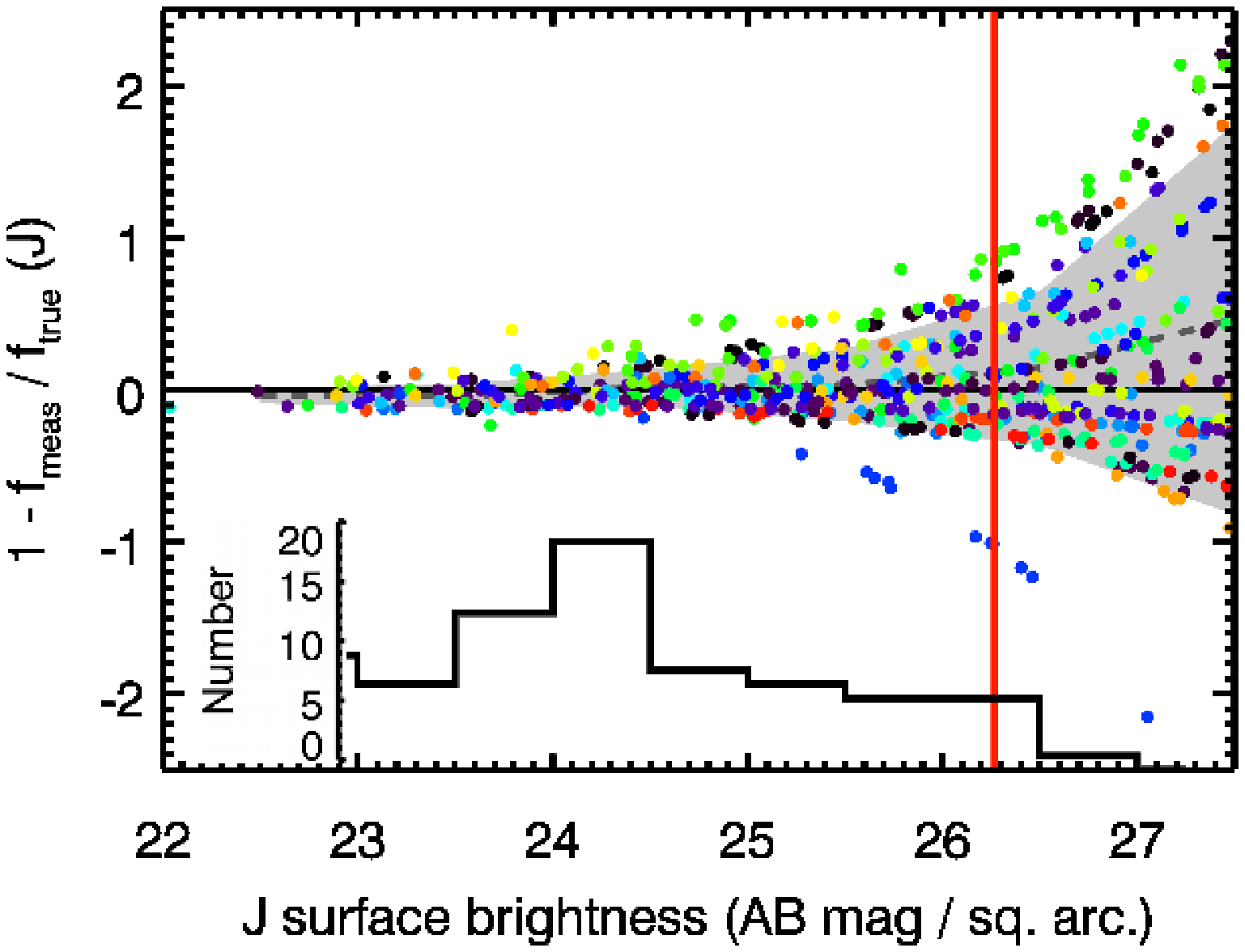}
    \includegraphics[width = 2.3in, height = 1.6in]{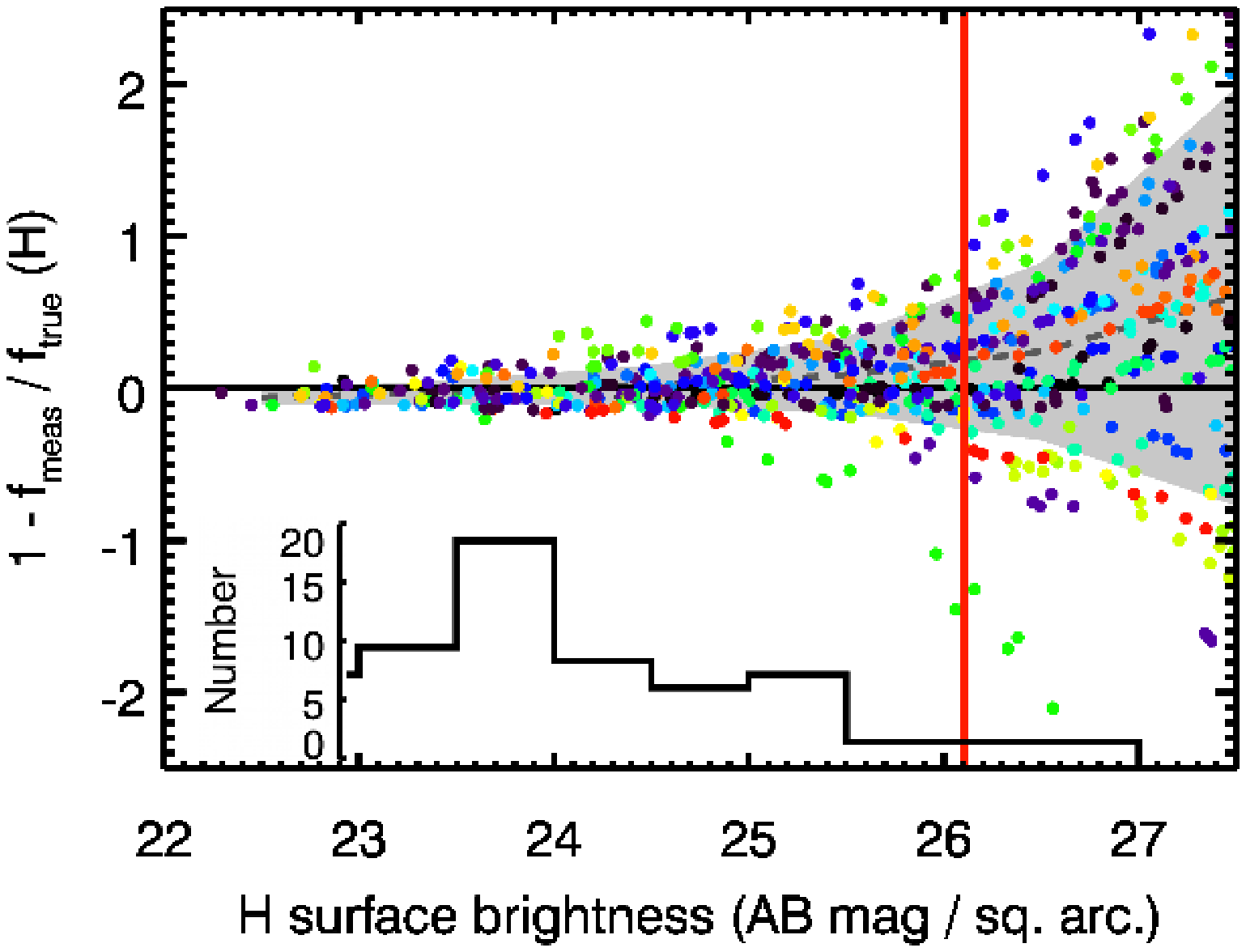}
    \includegraphics[width = 2.3in, height = 1.6in]{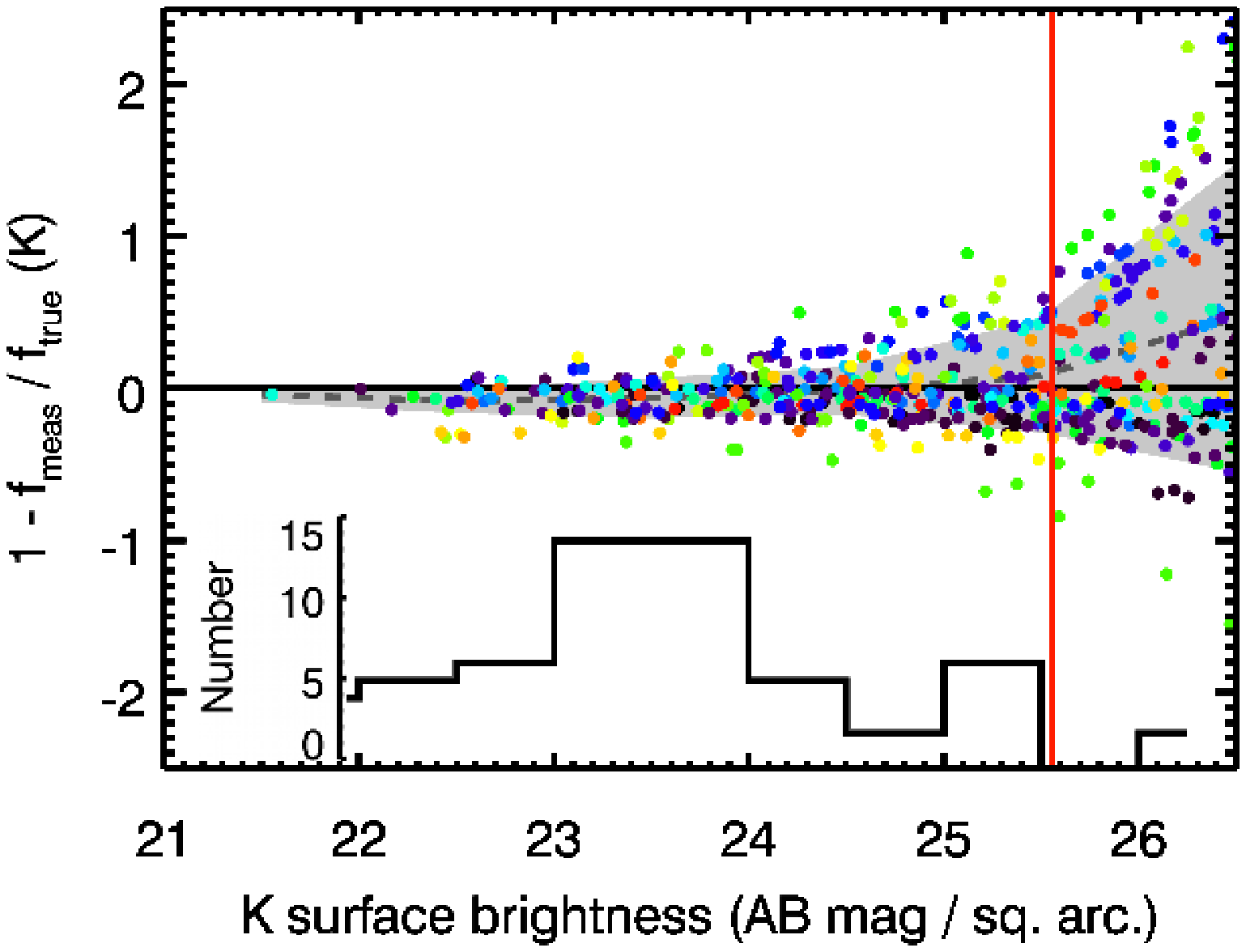}
 \end{center}
\end{figure}

Figure \ref{fig:flux_dif} plots the difference between the true surface brightness and measured surface brightness as a function of true surface brightness for 800 cloned galaxies in all observed bands.  The differences are plotted in relative flux units, or $(f_{meas} - f_{true}) / f_{meas}.$  Shaded regions denote the $1\sigma$ random error inferred by fitting Gaussian functions to the relative flux difference distributions.  The red vertical lines denote the surface brightness limit where the implied random photometric error exceeds $0.4$ magnitudes.  We choose this limit for two reasons:  First, the expected number of sources with negative flux measurement (a catastrophic failure) is limited to a few percent of the subsample whose random errors are of order $\sim$40\% in flux units.  Second, $0.4$ magnitudes of random error will average down to our expected level of systematic error ($\sim0.1$ mag) in samples with $\sim16-32$ sources.  

Histograms in all panels of Figure \ref{fig:flux_dif} show the distribution of measured surface brightnesses in each band for the ``all'' sample.  These histograms indicate that only a few sources are fainter than the limit corresponding to $0.4$ magnitudes of error (red line) in all observed bands redder than $B$.  In the $B$-band panel, a blue histogram denotes the distribution of $B$ surface brightnesses for sources in the ``all'' subsample with $z < 0.638$.  Sources below this redshift require the $B$-band to interpolate a rest-frame $u$ flux; sources above this redshift do not require $B$-band measurements.  Of all the observed bands, the highest photometric errors are seen in the $B$-band for the ``all'' sample.  However, none of the sources for which $B$-band is necessary to compute a rest-frame $u$ flux have a photometric error greater than $0.4$ magnitudes.

We conclude that, for the ``all'' sample as selected, rest-frame $u-R$ color can be reliably constrained for all sources that lack central point sources.

\subsubsection{Galaxies with Central Point Sources}

It is expected that the presence of central point sources will cause error in surface brightness measurements, although the use of a wide annulus (inner radius $0\farcs75$) will reduce this effect.  As nonthermal AGN emission tends to be bluer (rest-frame $u-R \sim 0.3 $) compared to galaxy light (rest-frame $u-R \sim 1-2 $), this error is likely to be larger in the observed $B$ and $V$-band for a sample with $0.5 < z < 1.5$.  We now quantify the photometric error in each band as a function of central point source strength through our cloning analysis.

For this analysis, the cloning simulation of section \ref{sect:sim_no_PS} is repeated, but with artificial point sources added of varying amplitudes.  The ratio of the total amplitude of the point source to the total ampltitude of the galaxy (measured over a $3\farcs$ diameter circle), or Point Source Fraction, is varied between 0.1 and 1000.  Aperture photometry is performed on each of the cloned galaxies and corrected for contamination as described in Section \ref{sect:apcorr}.   We then compare the corrected surface brightnesses to the intrinsic host surface brightnesses to evaluate photometric error as a function of Point Source Fraction. 

\begin{figure}
  \caption{Plots of relative flux differences $(f_{meas} - f_{true}) / f_{meas}$ in ACS and ISAAC bands for cloned galaxies with central point sources, plotted as a function of the Point Source Fraction.  Each panel plots simulated points for a different observed band.  All symbols are defined as in Figure \ref{fig:flux_dif}.  }
  \label{fig:PS_dif}
  \begin{center}
    \includegraphics[width = 2.3in, height = 1.7in]{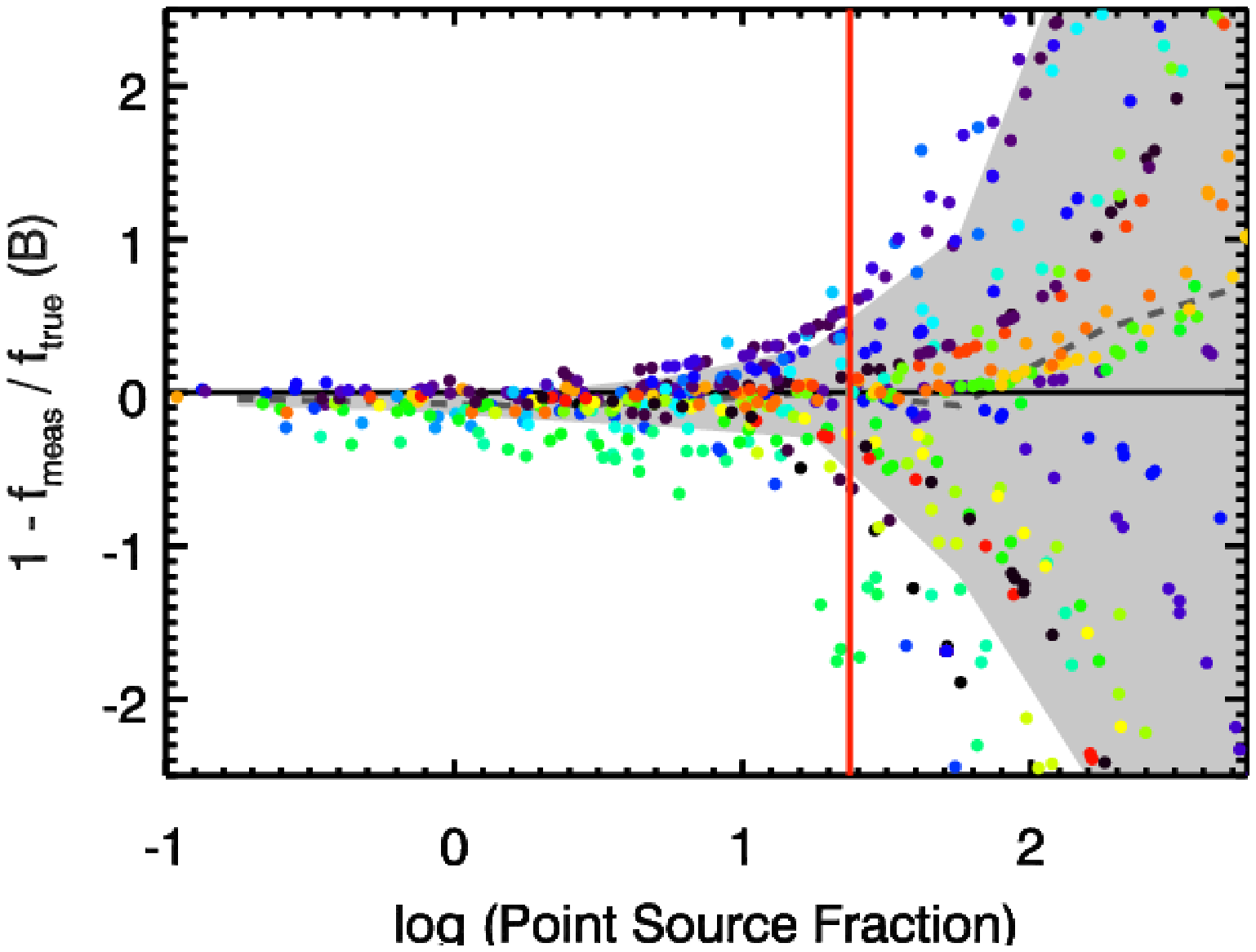}
    \includegraphics[width = 2.3in, height = 1.7in]{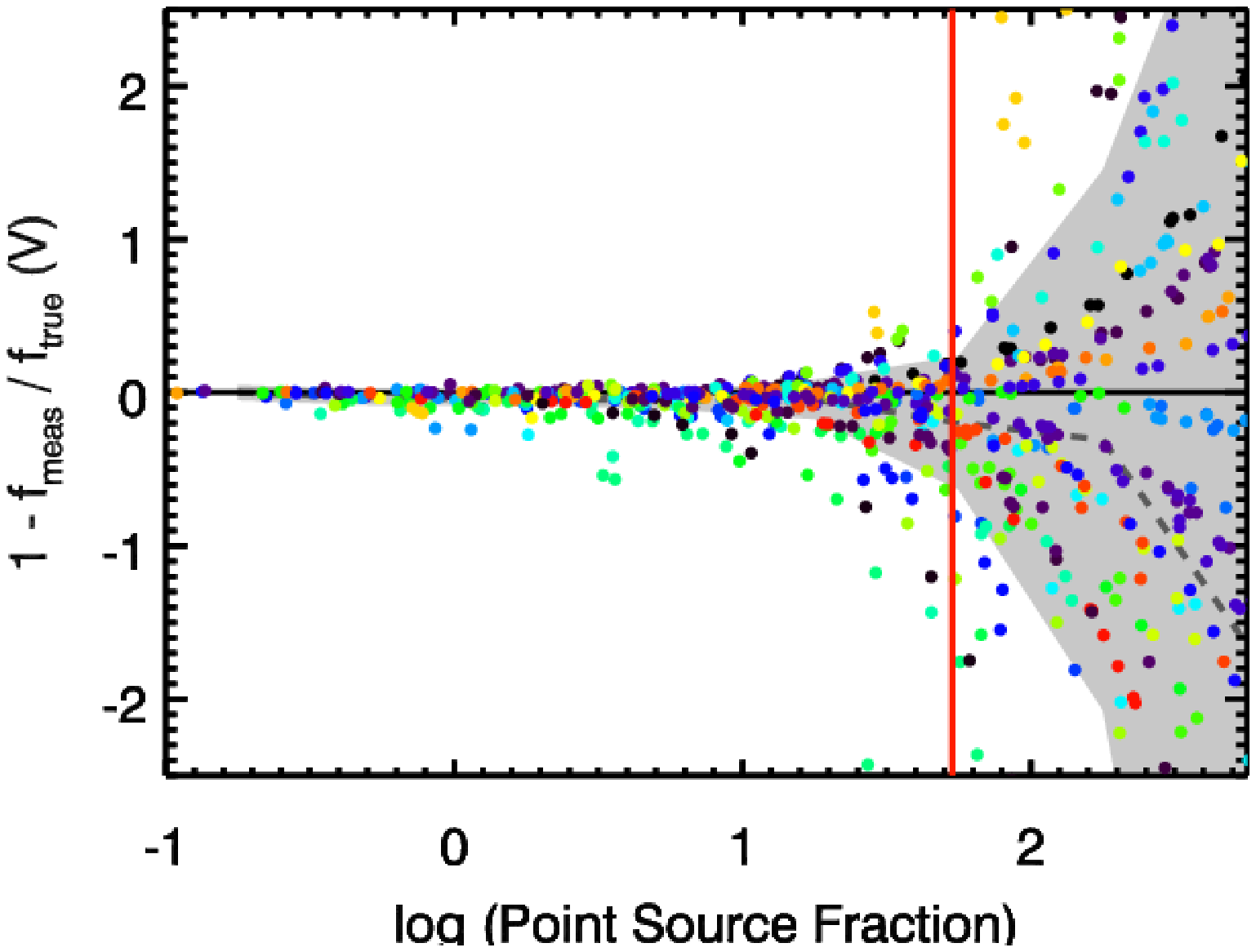}
    \includegraphics[width = 2.3in, height = 1.7in]{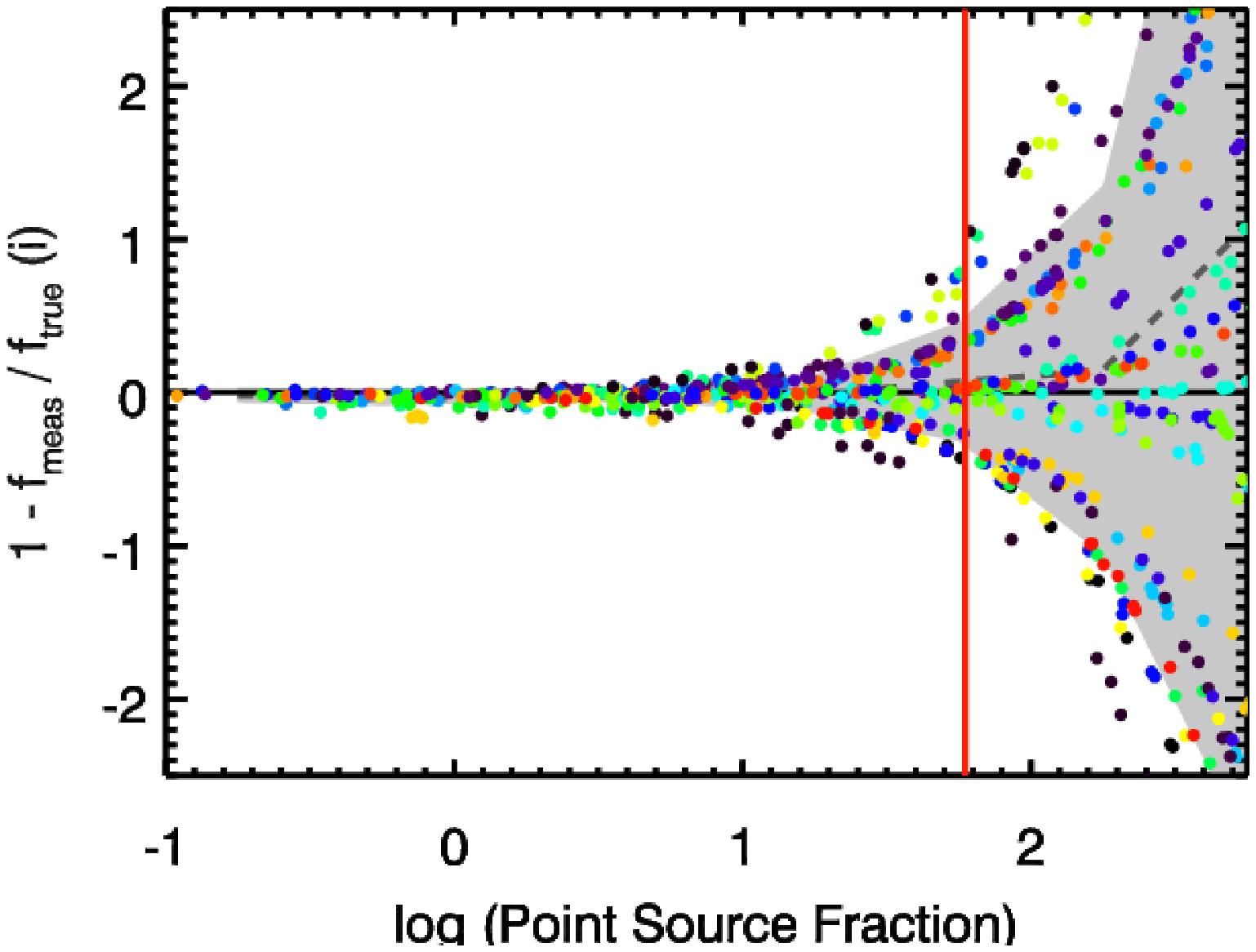}
    \includegraphics[width = 2.3in, height = 1.7in]{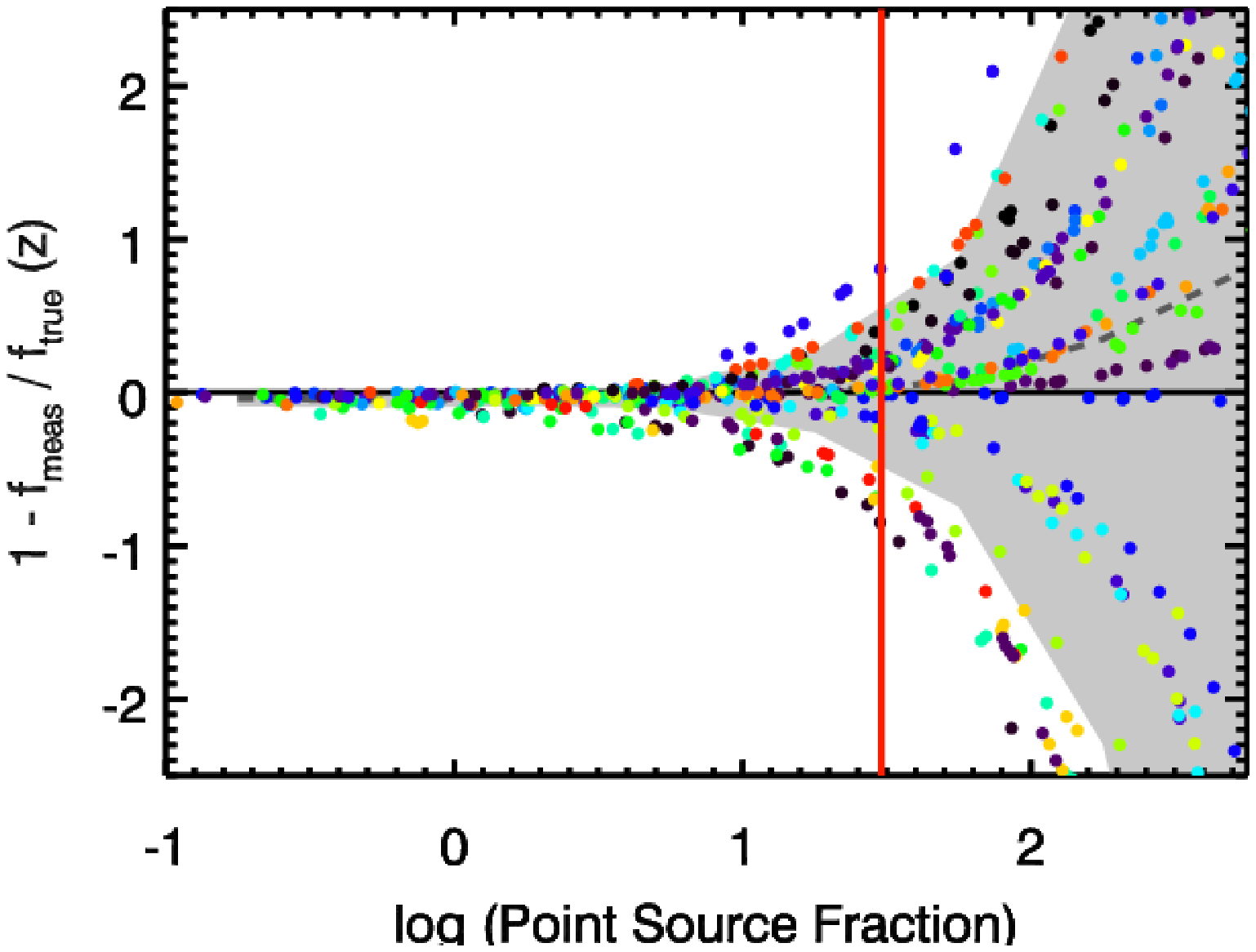}
    \includegraphics[width = 2.3in, height = 1.7in]{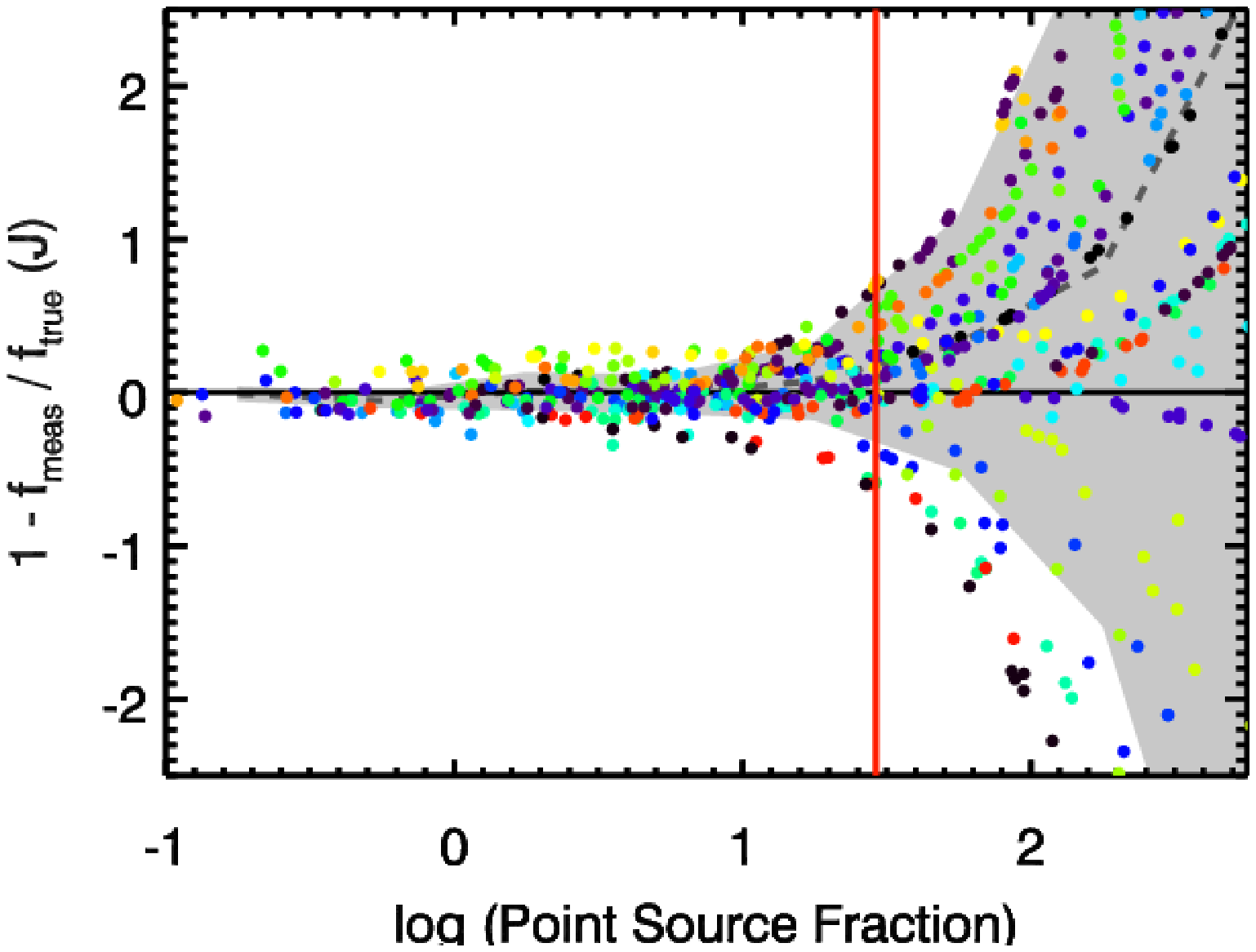}
    \includegraphics[width = 2.3in, height = 1.7in]{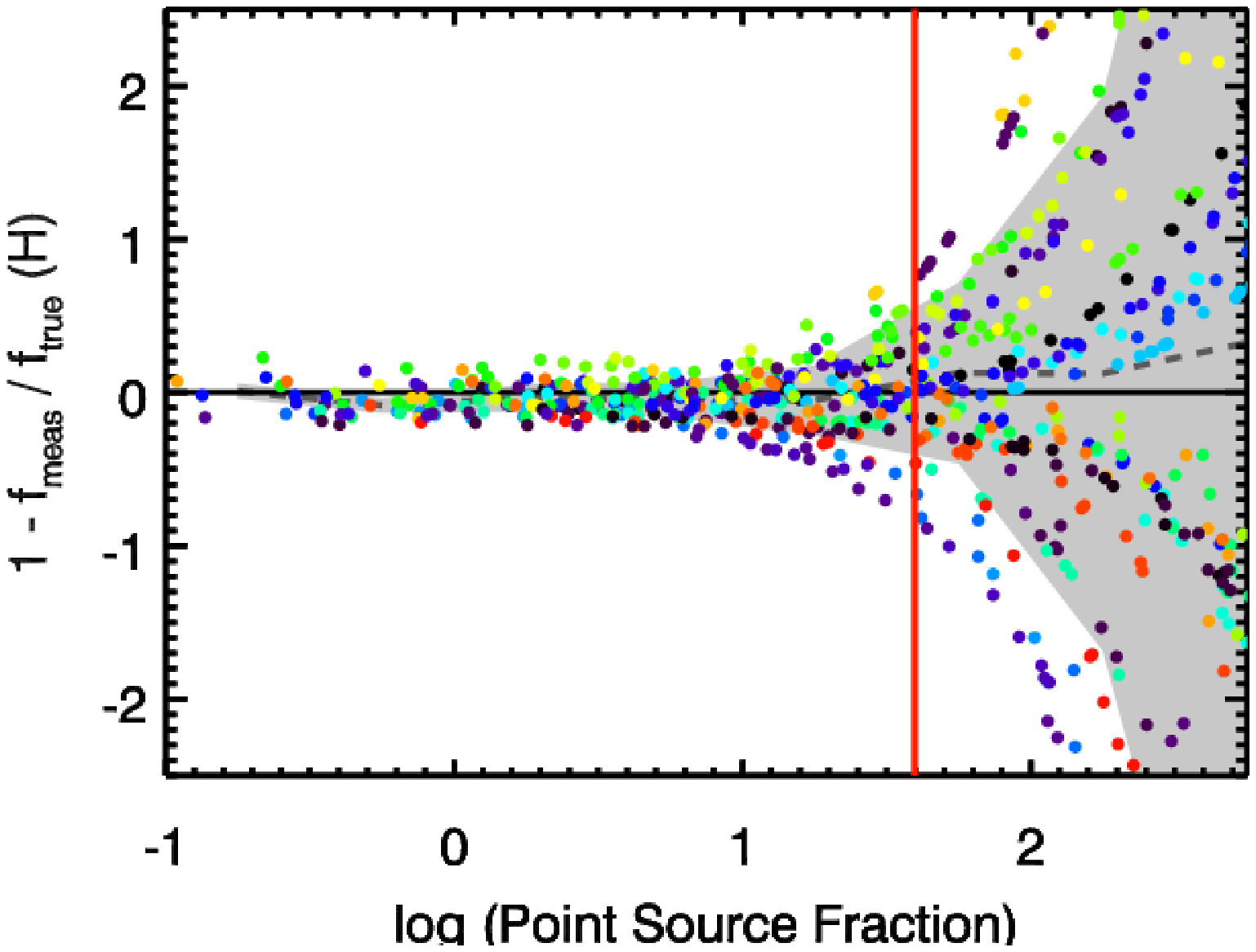}
    \includegraphics[width = 2.3in, height = 1.7in]{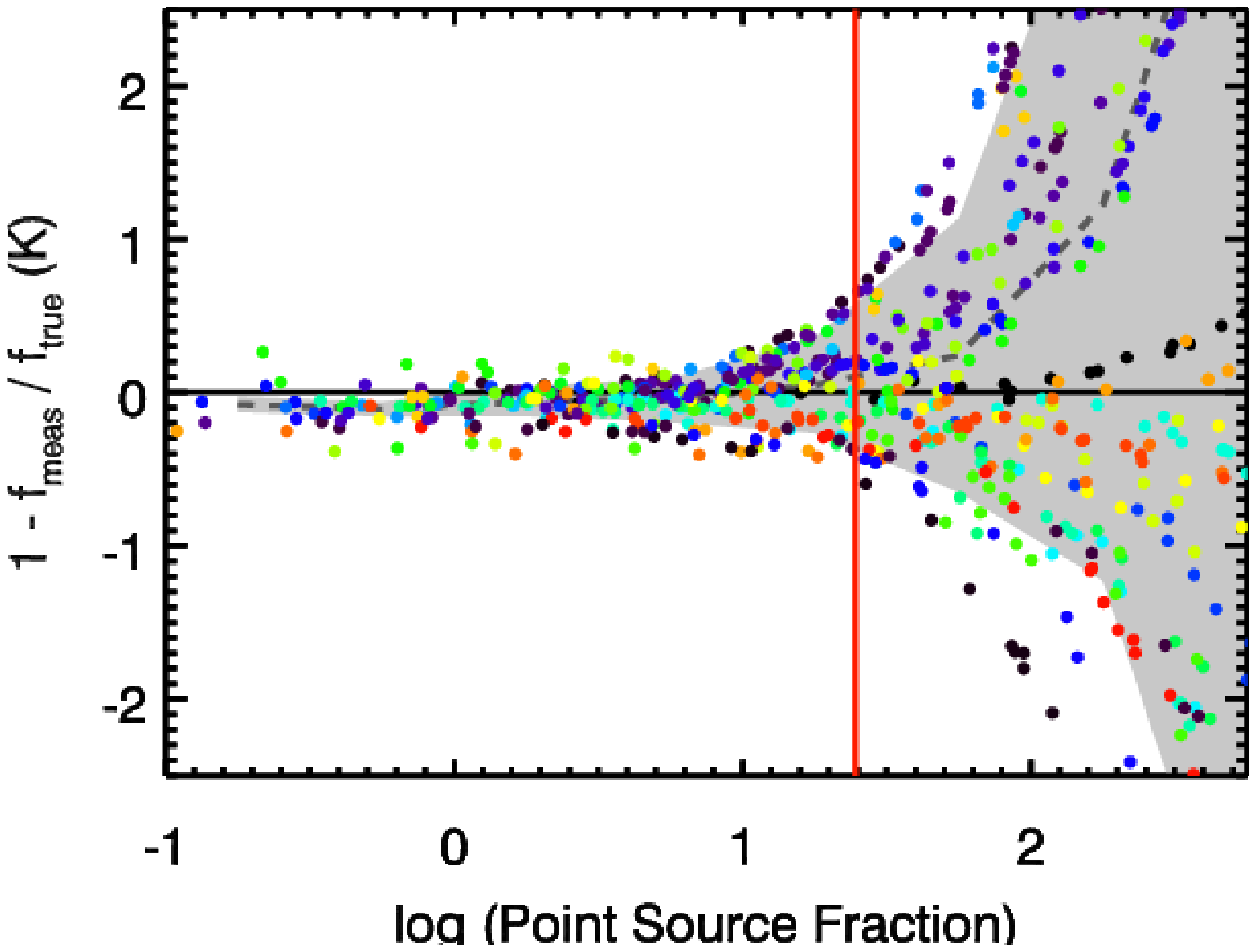}            
 \end{center}
\end{figure}

The relative flux differences in all bands are shown in Figure \ref{fig:PS_dif}, as in Figure \ref{fig:flux_dif}, but now plotted against the logarithm of the Point Source Fraction.  Shaded regions denote the $1\sigma$ random error inferred by fitting Gaussian functions to the relative flux difference distributions.  The red vertical lines denote the Point Source Fraction at which the random photometric error exceeds $0.4$ magnitudes.  There are clear trends between the reconstructed surface brightness and the amplitude of the central point source in all bands.  The photometric error is less than $0.4$ magnitudes for sources with Point Source Fraction less than $\sim25-50$ in each band.  Notice that the HST $V$ and $i$ bands have the least photometric error for a given Point Source Fraction and the $B$ and $K$ bands have the most. 

Figure \ref{fig:PS_dif} indicates that ``outer'' aperture flux measurements will be reliable for sources with Point Source Fraction less than 30 in all bands.  While the Point Source Fraction is difficult to measure for non-simulated data, a more easily-measured proxy can be used.  Here we use the ratio between the amplitude in a $0\farcs18$ aperture and a circular, annular aperture of inner radius $0\farcs18$ and outer radius $1\farcs$, or $C_{0.18/1.0}$.  We define the Normalized Concentration $C_{0.18/1.0, norm}$ as 

$$ C_{0.18/1.0, norm} = \frac{C_{0.18/1.0, image}}{C_{0.18/1.0, PSF}}. $$

\begin{figure}
  \caption{Plot of measured log($C_{0.18/1.0, norm}$) as a function of input Point Source Fraction for the 800 cloned galaxies in this simulation.  Dashed lines mark fitted spline functions.  The tightness of the correlation suggests that the easily measured $C_{0.18/1.0}$ is a good proxy for point source contamination.  }
  \label{fig:conc_PS}
  \begin{center}
    \includegraphics[width = 5.4in, height = 4.1in]{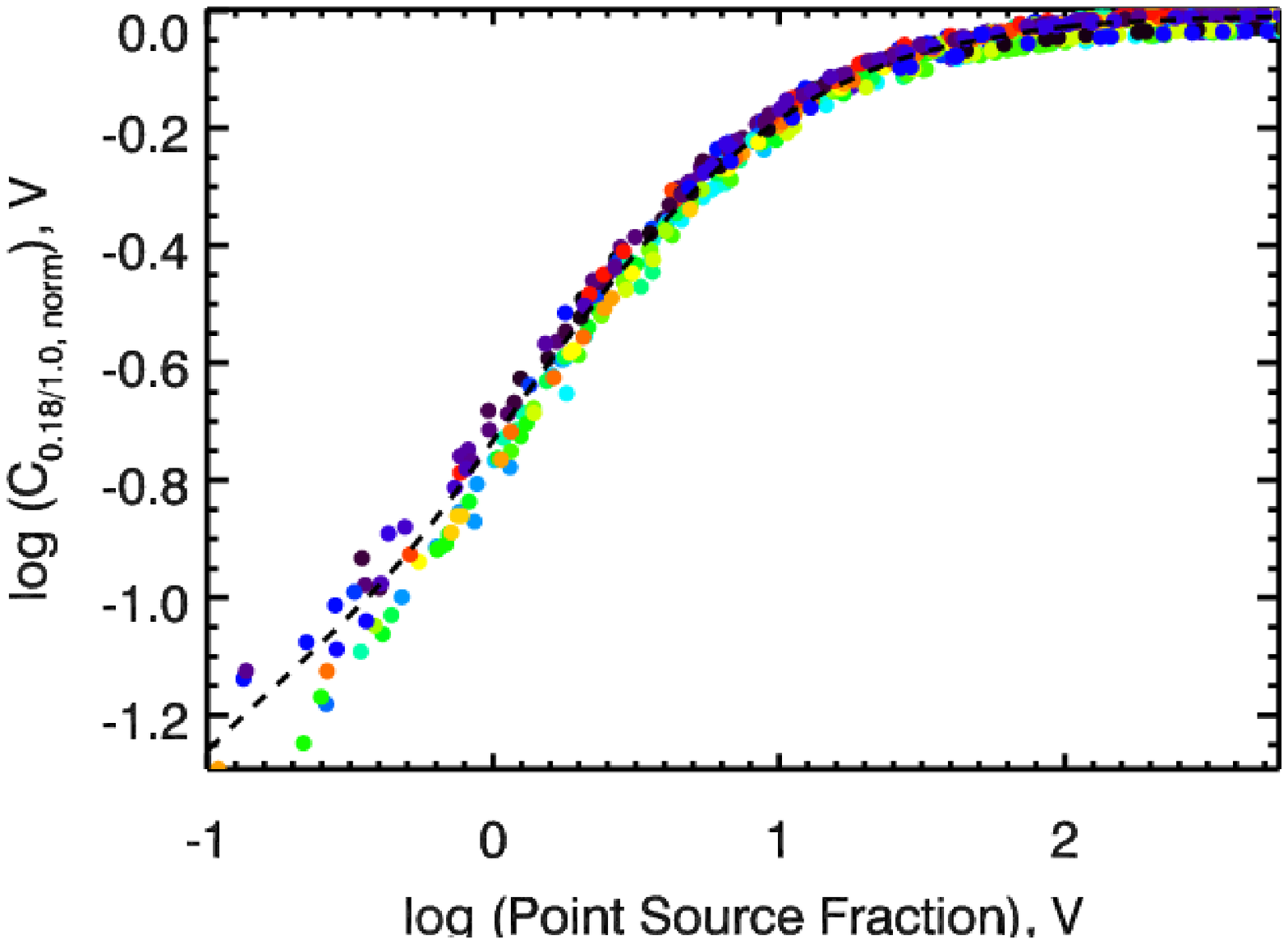}

 \end{center}
\end{figure}

This parameter will correct for PSF variation in a field.  Figure \ref{fig:conc_PS} plots the relationship between $C_{0.18/1.0, norm}$ and the known Point Source Fraction as measured for the cloned galaxies above in observed $V$.  The tightness of the correlation suggests that measurements of $C_{0.18/1.0, norm}$ in actual galaxies can be used to look up the Point Source Fraction for a given band.

\subsection{Correction for Contamination in Integrated Fluxes}
\label{sect:corr_contam_int}

Constructing CMDs that are corrected for point source contamination requires an unbiased estimate of the host galaxy's absolute magnitude.  In practice, this is more difficult than estimating uncontaminated color in the ``outer'' aperture, as it requires measurement of the integrated galaxy flux beneath a central point source.  Here, we describe our method of using imaging to estimate point source contamination for a given band.

The tightness of the correlation in Figure \ref{fig:conc_PS} suggests that $C_{0.18/1.0, norm}$ may be used as a proxy for point source contamination.  To estimate the point source contamination, we first measure $C_{0.18/1.0}$ in all bands.  Next, for each band, we fit separate spline functions to the $C_{0.18/1.0, norm}$ / Point Source Fraction relations in the simulated galaxies (e.g., Figure \ref{fig:conc_PS}).  These spline functions are inverted to create monotonic look-up tables.  Finally, the $C_{0.18/1.0, norm}$ values are used to look up the Point Source Fraction in each band.  The Point Source Fraction is used to calculate and subtract a correcting factor from the measured integrated photometry.

We estimate rest-frame $R$-band absolute magnitudes by performing this procedure in the observed $JHK$ bands.  These are displayed in the next section in corrected CMDs for the ``all'' sample.

\subsection{Implications for Observations}

Our cloning studies have constrained systematic and random measurement errors as a function of intrinsic surface brightness and Point Source Fraction in all bands.  We use these relations to calculate systematic and random errors for each AGN host in the ``all'' sample.  We estimate Point Source Fraction as described in Section \ref{sect:corr_contam_int}.  We note that using {\it measured} instead of {\it intrinsic} surface brightnesses to look up errors will underestimate error bars for some of the faintest sources.  However, the original sample cut of observed $R < 24$ (total galaxy magnitude) has largely removed intrinsically faint sources from the sample, and very few of the remaining sources have photometric errors larger than $0.4$ magnitudes.  

\section{RESULTS}

\subsection{Corrected Color-Magnitude Diagrams}

\begin{figure}
  \caption{Rest frame $u-R$ vs. $M_R$ color-magnitude diagram, corrected for point source contamination.  Symbols are plotted as in Figure \ref{fig:UR_CMD}.  $u-R$ colors have been measured with annular ``outer'' photometric apertures.  $M_R$ magnitudes are measured with integrated apertures.  Sources judged to be marginally contaminated by a central point source associated with AGN emission (Point Source Fraction in observed B greater than 0.35) are corrected for this contamination in $M_R$.  For these contaminated sources, uncorrected, integrated measurements are shown as faint, empty points.  Uncorrected, integrated loci are connected to corrected, ``outer'' loci with dashed lines.  Rest-frame values are estimated from observed photometry as in the text.  Blue and red error bars denote the mean colors and luminosities for the soft and hard sources, respectively, with errors calculated as in Section \ref{sect:error}.   Small black circles denote the colors and magnitudes of the underlying galaxy population, selected as explained in Section \ref{sect:CMD}.  Note that photometry is measured with $3\farcs$ filled apertures for this control sample, not ``outer'' apertures.  The soft source at the bluest extreme of the diagram (XID 100) has a significantly bluer ``outer'' color than its integrated color, unlike the remainder of the ``all'' sample; it appears to be composed of a blue point source with a bluer star-forming extension to the NE.}
  \label{fig:UR_CMD_vectors}
  \begin{center}
    \includegraphics[width = 6.0in, height = 4.3in]{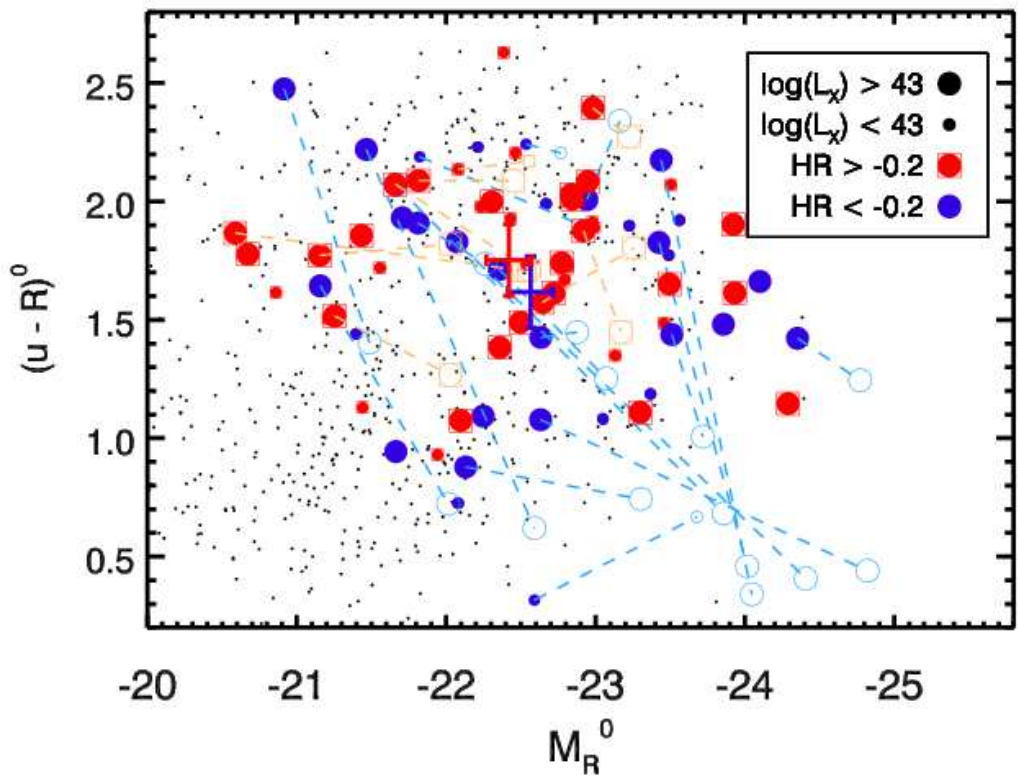}
 \end{center}
\end{figure}

Figure \ref{fig:UR_CMD_vectors} plot ``outer'' $u-R$ colors against corrected $M_R$ for the ``all'' AGN sample.  Rest-frame $M_R$ is estimated from observed $JHK$ integrated photometry ($3\farcs0$ filled aperture) via interpolation.  Corrections for point source contamination are measured using source concentration as described in Section \ref{sect:corr_contam_int} above.  Figure \ref{fig:UR_CMD_vectors} shows uncorrected $u-R$ and $M_R$ as light blue and light red points.  These uncorrected loci are linked with the corrected photometry with vectors. 

The simulations above have shown that photometric measurements using annular apertures avoid color contamination due to central point sources reliably.  However, estimating uncontaminated absolute magnitude is more difficult, as it involves estimating the flux directly beneath the AGN point source.   For large point source contamination (Point Source Fraction $>$ 30), this method produces uncertain results.  Therefore, we make no conclusions about the AGN samples involving the mean or individual values of $M_R.$  

It is clear in Figure \ref{fig:UR_CMD_vectors} that the difference in mean rest-frame $u-R$ color between soft and hard sources seen in Figure \ref{fig:UR_CMD} is drastically reduced by using annular photometric apertures and correcting for contamination from AGN emission.  The mean colors are also listed in Table \ref{tab:tabulated_results}; note that the use of ``outer'' apertures reduces the soft-hard color difference by a factor of nearly three.  Given the measurement errors, the mean rest-frame $u-R$ color of the ``all-soft'' and ``all-hard'' subsamples are statistically equivalent.

Note that the mean color and magnitude of the obscured population is different in Figure \ref{fig:UR_CMD} and \ref{fig:UR_CMD_vectors}, despite the lack of point sources in this population.  The corrected ``outer'' magnitudes of the obscured population are $\sim0.1$ bluer than the uncorrected, integrated magnitudes.  Because of the negative color gradients in the host galaxies, the outer regions are bluer than the nuclear regions.  For the unobscured sources, this effect is overwhelmed by the removal of the central point source contamination, which causes the corrected ``outer'' magnitudes to be redder than the integrated magnitudes.

\subsection{Mean SED Comparison}

\begin{figure}
  \caption{Logarithmic mean of surface brightnesses for all-soft sample and all-hard sample.  The dashed lines represent the one-sigma uncertainties on the mean SEDs.  Red lines denote measurements for soft AGN and blue lines denote measurements for hard AGN.  Error bars are measured from full cloning simulations and a bootstrap procedure to estimate error due to finite sampling.  SEDs of \citet{bru03} Single Stellar Population (SSP) bursts of varying ages are shown for comparison.  }
  \label{fig:type_comp}
  \begin{center}
    \includegraphics[width = 5.3in, height = 4.6in]{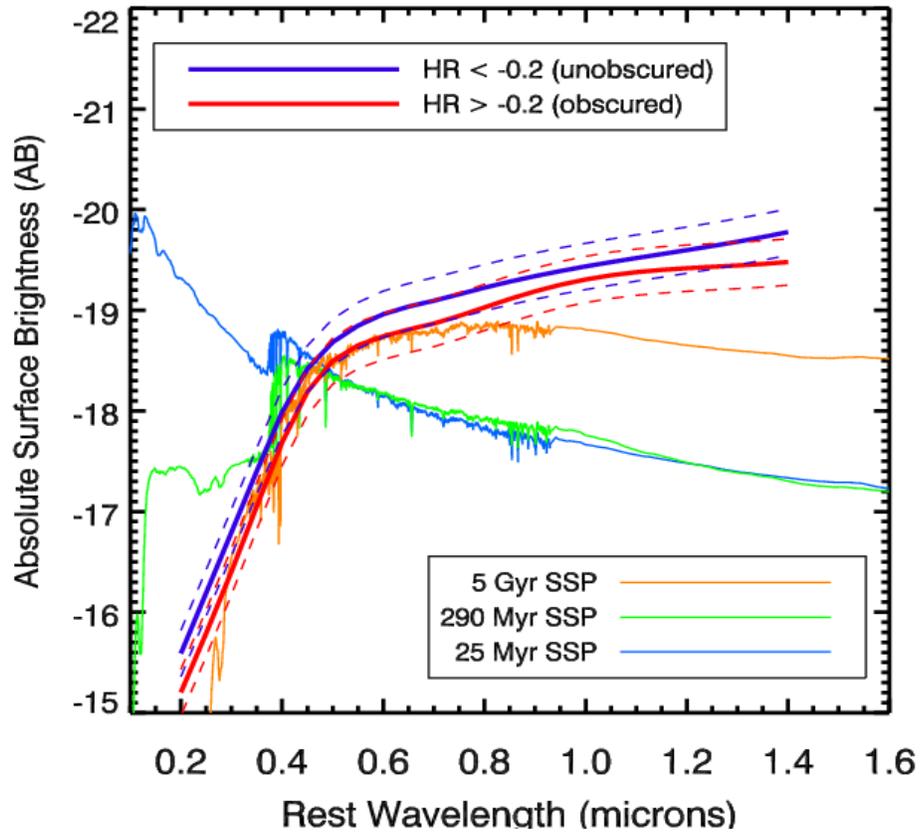}

 \end{center}
\end{figure}

The surface brightnesses of the AGN sample are plotted as de-redshifted spectral energy distributions (SED) in Figure \ref{fig:type_comp}.  Surface brightnesses are plotted in absolute AB magnitudes, modified to account for a de-projection of inclined disks (i.e., fluxes are multiplied by $b/a$, where $b$ is the semiminor axis and $a$ is the semimajor axis).  Figure \ref{fig:type_comp} plots the logarithmic mean (or geometric mean of magnitudes) of these SEDs for the ``all-soft'' and ``all-hard'' subsamples.  SEDs of \citet{bru03} Single Stellar Population (SSP) bursts are shown for comparison.  The SEDs of the soft and hard sources appear to be similar, although the soft sources are $\sim0.3$ magnitudes brighter than the hard sources at all wavelengths. 

The mean SEDs of the two populations are inconsistent with a starburst population of age $\sim25$ Myr.   The SEDs more resemble those of intermediate and old stellar populations, although the rise at near-infrared wavelengths compared to these SSP models implies that dust extinction is present.  Young stellar populations and their accompanying ultraviolet emission may be hidden by this dust.

\subsection{Mean Color Gradients in ERS sample}
\label{sect:ERS_colors}
The principal advantage of using the ``outer'' aperture is the avoidance of central point sources; the primary disadvantage is that is does not probe the central regions of host galaxies.  The inner radius of the ``outer'' aperture was set by the image quality of the ISAAC JHK imaging, approximately $0\farcs65$ in the worst case.  We now utilize $F125W$ and $F160W$ imaging in GOODS-South from the WFC3/IR imager aboard HST, with a spatial resolution of $\sim0\farcs2$, to constrain host colors at smaller radii.  Although the spatial coverage of the WFC3 infrared imaging is less than 50 square arcminutes, the improved depth relative to ISAAC ($H_{AB} = 27.0$ for point sources, $5\sigma$) permits extending the sample to fainter magnitudes.

\begin{figure}
  \caption{Mean color gradients for three populations in the WFC3 ERS sample: Obscured
sources (HR $> -0.2$), all unobscured sources (HR $< -0.2$), and only unobscured sources
with little point source contamination (V Point Source Fraction $<$ 10). Filled regions are 68\%
confidence intervals. With the removal of sources strongly contaminated by central
point sources, the mean color gradients of the obscured and unobscured sources
are similar to $\sim0.5$ magnitudes at all radii greater than 1 kpc.}
  \label{fig:color_grad}
  \begin{center}
    \includegraphics[width = 5.5in, height = 4.3in]{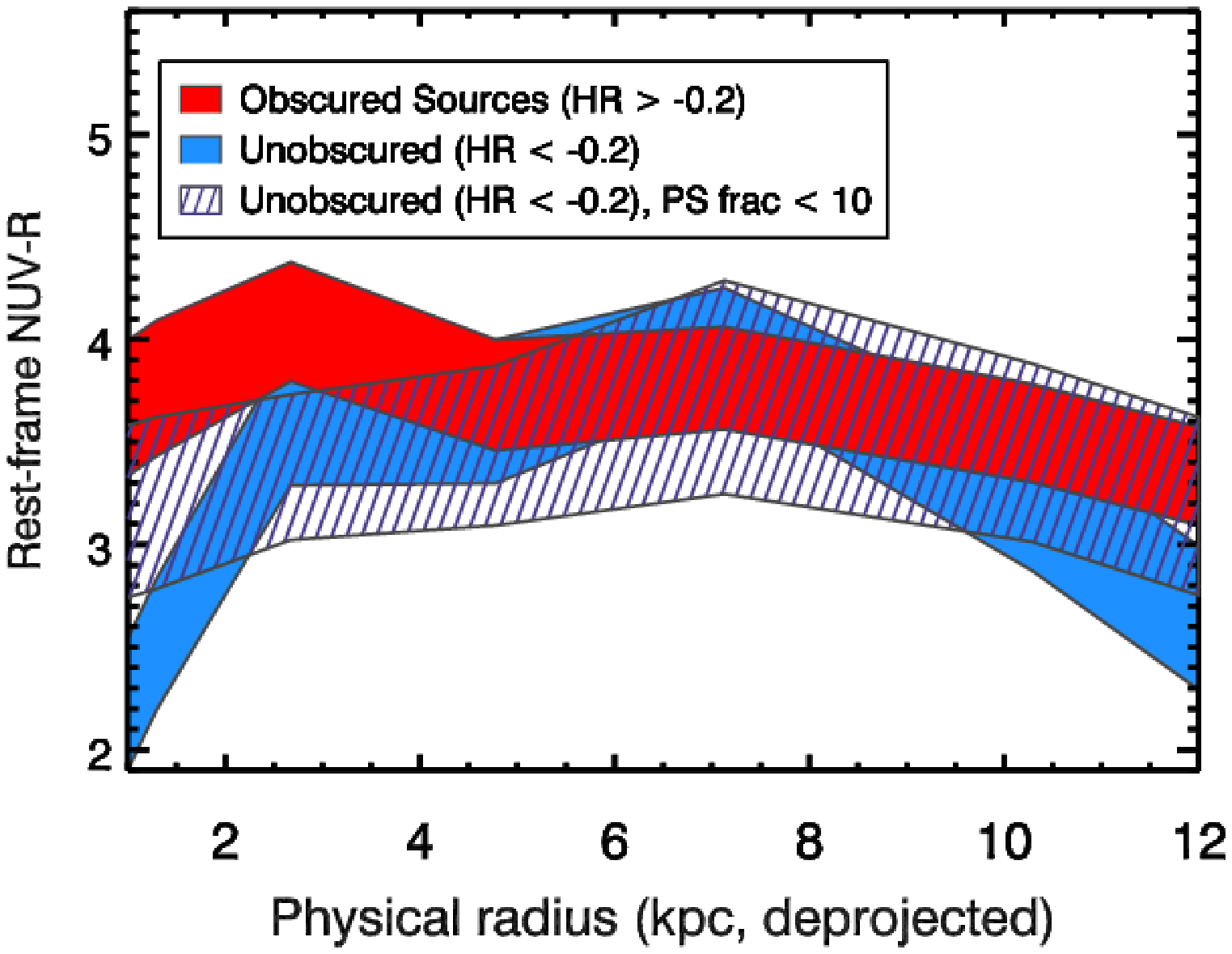}

 \end{center}
\end{figure}

We compute $NUV-R$ color gradients for each host galaxy as described in Section \ref{sect:color_grad_method}.  Figure \ref{fig:color_grad} displays the mean $NUV-R$ color gradients for the ``ERS-soft'' and ``ERS-hard'' subsamples.  The shaded regions denote the 68\% confidence intervals as computed in Section \ref{sect:error} below.  Two curves are displayed for soft sources; the blue curve shows the mean color gradient for all 15 soft sources and the hatched curve shows the mean color gradient for the 10 soft sources whose Point Source Fraction in V-band is less than 10.  Point Source Fraction is measured in the $V$-band as described in Section \ref{sect:corr_contam_int}.

Note that because the images are CLEANed and the central aperture is filled, contaminating point sources will affect the color of the innermost point only ($r = 1.3$ kpc) in Figure \ref{fig:color_grad}.  The innermost point of the mean color gradient of the full soft sample, denoted by the blue curve, is significantly bluer than all other points.  However, when sources with significant point source contamination (Point Source Fraction $>$ 10) are removed from the soft sample, as shown with the hatched curve, the innermost point returns to a color that is similar to the colors at all other radii.  The removal of these contaminated sources does not affect the mean colors of the hosts beyond $r \sim 2$ kpc.  

The colors of the hard and soft sources are statistically identical at all radii beyond $r \sim 2$ kpc whether objects with significant point source contamination have been removed or not.  Removing objects with central point sources from the soft sample allows us to make an unbiased color comparison between soft and hard sources for the central $r=1.3$ kpc point; when these objects are removed, the remaining soft sample and the hard sample are statistically equivalent in color at $r=1.3$ kpc.  Assuming that the point source contamination in the objects with Point Source Fraction $>$ 10 is caused by nonstellar emission associated with the AGN, we may conclude that the colors of the soft and hard sources for this subsample are similar at all radii greater than $r\sim1$ kpc.  

\subsection{Error Considerations}
\label{sect:error}
The method of averaging can possibly affect the mean photometric properties of a sample.  When the averaging is performed in flux space, with equal weights on each photometric point, the principal outcomes of the comparisons between the various types of AGN in this sample are unchanged.  Similarly, when the averaging is performed in flux space with stronger weights towards fainter fluxes, the comparison is unchanged.

Systematic measurement errors are seen in the cloning simulation; we estimate these for individual galaxies and subtract them.  The random errors on the mean SEDs at a particular wavelength and on the mean colors of subsamples are the sum of two components: The first is a random error caused by sky-subtraction errors, photon error in the photometric measurement, and other errors associated with basic aperture photometry.  The second major error, dominant over photometric error, is due to small-number statistics in these sub-samples.  We estimate this error with a Monte Carlo bootstrap method, as outlined in \citet{pre92}.  In this method, new sub-samples are repeatedly drawn (with replacement) from the overall data set, analyzed, and used to populate a distribution.  The random error due to the finite size of the sample is then given by the standard deviation of critical parameters (e.g., mean rest-frame $u$ magnitude) in this distribution. 

The result of this analysis is that the bootstrap error in $u-R$ is 0.11 magnitudes for the all-soft subsample and 0.10 magnitudes for the all-hard subsample.  This can be understood in terms of the intrinsic brightness distribution and color scatter in the samples.  The intrinsic variation in $u-R$ color for this sample is $\sigma \sim0.7$ magnitudes.  The error on the mean of some parameter due to incomplete sampling of a population goes as $\sigma / \sqrt{N},$ where $\sigma$ is the standard deviation of the parameter for the parent population and $N$ is the size of the sample.  With $N \sim 31-42$ for the all-soft and all-hard subsamples, respectively, the error on the mean is $\sim 0.13$ and $\sim0.11$ magnitudes for $u-R$ color, respectively.  These alternate estimates are nearly the same as that estimated with a Monte Carlo bootstrap procedure.  

\subsection{Are Obscured AGN Hosts More Dusty than Unobscured AGN Hosts?}

\begin{figure}
  \caption{Top panels:  Rest-frame $U - V$ vs. $V - J$ color-color diagrams of all galaxies in a mid-IR selected sample with $L_V > 5 \times 10^9 L_{\odot}$ (included with permission from \citet{wuy07}).  SDSS+2MASS galaxies (small gray dots) are plotted as a local reference.  Top left panel:  Galaxies are color-coded by dust extinction.  Dust extinction affects rest-frame $V - J$ color more than $U - V$.  Top right panel:   Galaxies are color-coded by mean stellar age.   Stellar age affects rest-frame $U - V$ color more than $V - J.$  Bottom panel:  Rest-frame $U - V$ vs. $V - J$ color-color diagram using ``outer'' aperture for AGN with observed $Ks$ available.  Small black circles denote the colors of the underlying galaxy population, selected as explained in Section \ref{sect:CMD}.  Note that photometry is measured with $3\farcs$ filled apertures for this control sample, not ``outer'' apertures.  The mean colors of the AGN sample are shown in the center of the diagram; the obscured locus is close to the unobscured locus in $V - J$, suggesting that the populations have similar mean extinctions.}
  \label{fig:dust_plots}
  \begin{center}
    \includegraphics[width = 2.3in, height = 2.2in]{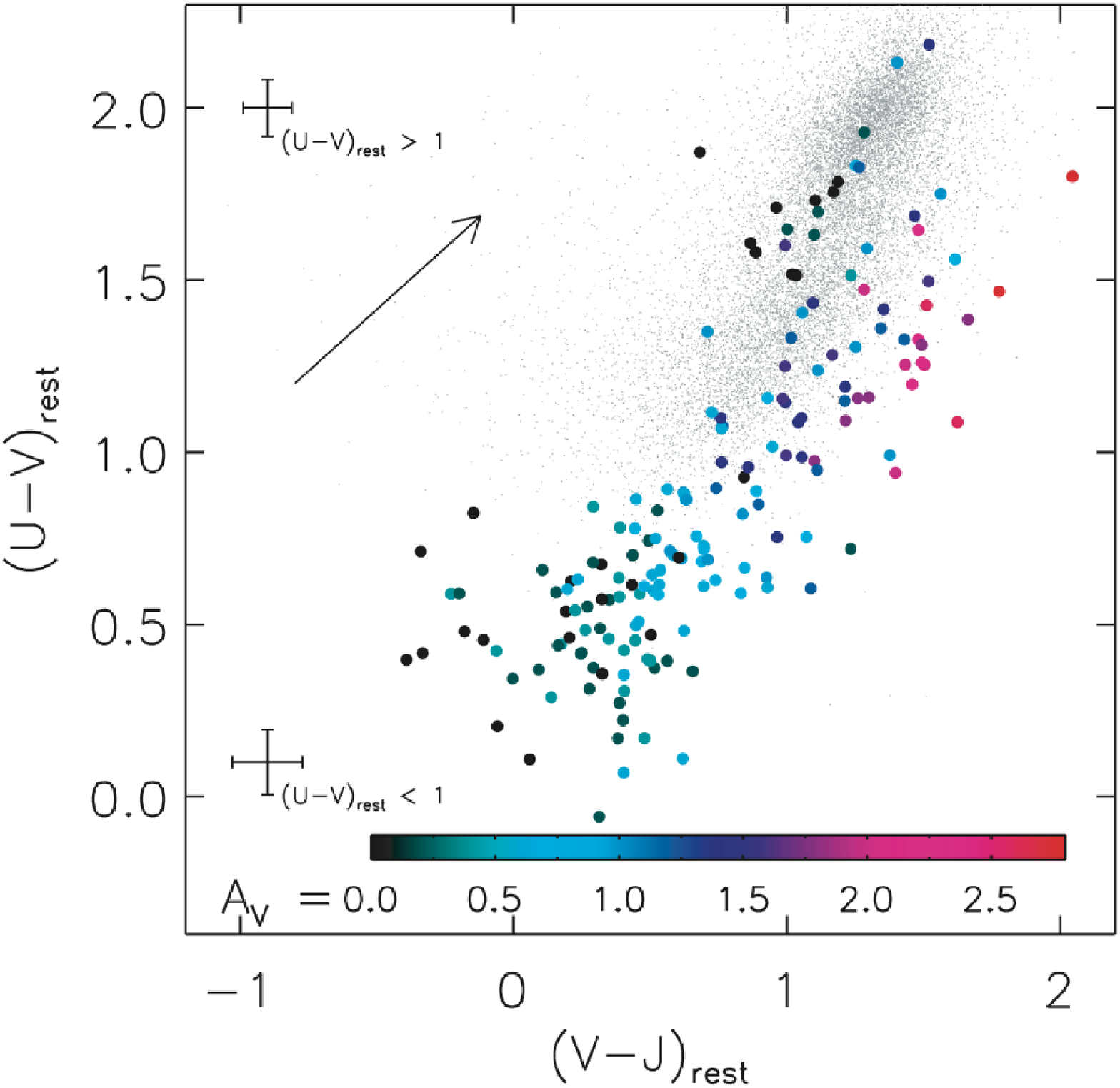}
    \includegraphics[width = 2.3in, height = 2.2in]{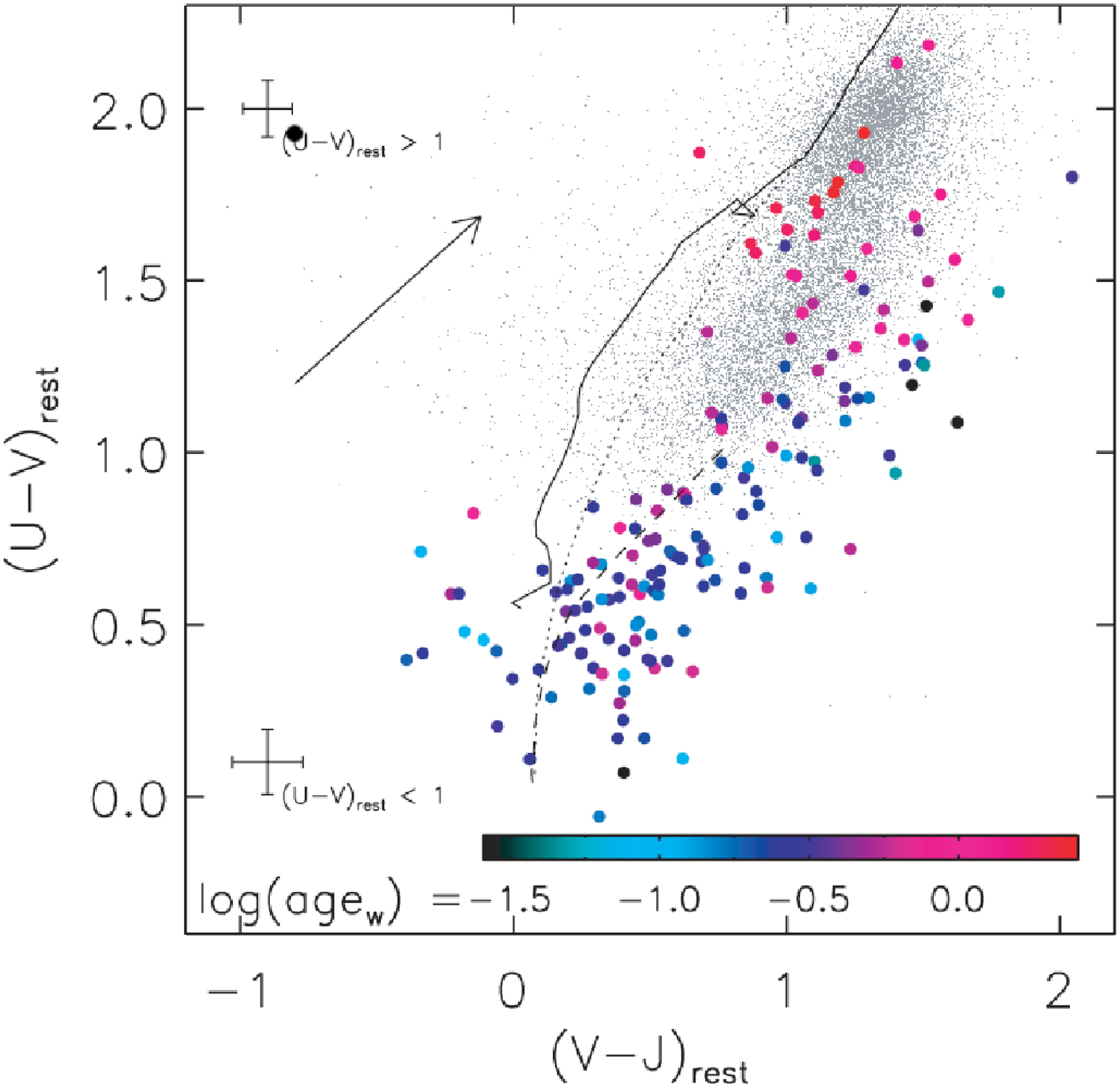}
    \includegraphics[width = 4.8in, height = 4.3in]{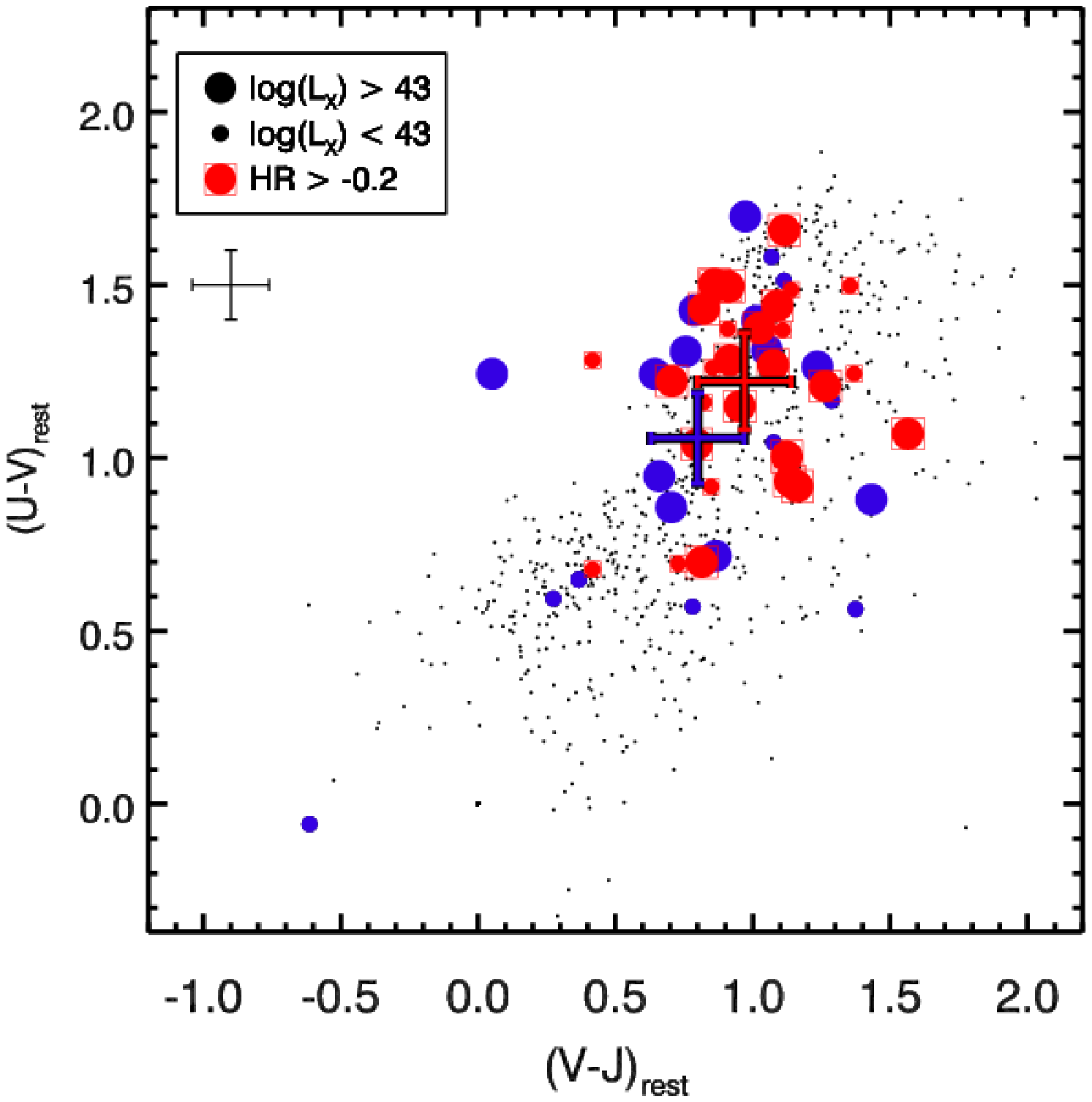}
 \end{center}
\end{figure}

$u - R$ and $NUV-R$ colors are sensitive probes of low levels of star formation \citep{sal07}, but these colors can be affected by dust extinction. Here we search for systematic differences in galactic dust extinction between the obscured and unobscured AGN host populations.  In a rest-frame $U-V$, $V-J$ color-color diagram, the effects of dust extinction can be separated to some extent from the effects of stellar age (see Figure \ref{fig:dust_plots}).  $V-J$ color is more sensitive to dust extinction than stellar age \citep{wuy07}. 

We compute $V-J$ ``outer'' colors to estimate the differential effects of dust extinction.  Rest-frame $J$ band magnitudes are calculated from observed \textit{BVizJHK} photometry using linear interpolations as described in Section \ref{sect:apphot}.  Half of the sample has $z > 0.85$, beyond which rest-frame $J$ shifts redder than observed $Ks$, the reddest band available in this sample.  Color extrapolations are required beyond this redshift.  We estimate the error in this extrapolation by measuring the natural dispersion in rest-frame $z-J$ ``outer'' color (observed $H-Ks$) at a single rest-frame $R-z$ (observed $J-H$) color slice for sources with $z < 0.85$.  This dispersion is 0.10 magnitudes, implying that the extrapolation error is less than 0.1 magnitudes for all sources in the sample.  The error on the mean of the samples will be much smaller than this.

The bottom panel of Figure \ref{fig:dust_plots} plots the ``outer'' colors of the ``all'' sample in a rest-frame $U-V$, $V-J$ color-color diagram.  Only sources for which observed $Ks$ is measured are included, i.e., only sources with ISAAC imaging in the southern GOODS field.  The mean rest-frame $U-V$ color of the two populations are discrepant by 0.16 magnitudes, which is consistent with the results shown in Table \ref{tab:tabulated_results} within the errors.  The mean $V-J$ color of the obscured population is redder than the unobscured population by $0.17$ magnitudes.  These differences are not statistically significant, suggesting that the two populations are similar in mean galactic dust extinction.  The two populations do not occupy clearly separated regions in the diagram.  Approximately one-half of the sample falls in a region corresponding to $A_V > 1$, suggesting that a significant fraction of the sources are dusty, which is consistent with other studies using similar data and methodology \citep{car10}.  

\subsection{$u-R$ Properties of AGN Hosts}

As seen in Figure \ref{fig:type_comp} and Table \ref{tab:tabulated_results}, the mean SEDs and mean rest-frame $u-R$ colors of AGN hosts beyond $r \sim 6$ kpc are largely similar across X-ray obscuration.   The hosts of the soft AGN, on average, are $\sim0.11$ magnitudes bluer in $u-R$ and $\sim0.2$ magnitudes bluer in $NUV-R$ than the hosts of the hard AGN beyond $r \sim 6$ kpc.  This effect is weak and not statistically significant.  When this analysis is repeated for luminous objects with $L_X > 10^{43}$ ergs s$^{-1}$, the $u-R$ color of the soft sources is statistically equivalent to that of the hard sources.  The soft sources are $0.17$ magnitudes bluer than the hard sources, but this effect is not statistically significant.

We have checked this result with a simple stacking exercise, in which aperture photometry is performed on a stack of the processed images of a particular population.  For this procedure, we use sky-subtracted images whose ACS/ISAAC alignment is good to $\sim 0\farcs1$ that have been deconvolved with CLEAN as described in section \ref{sect:CLEAN}.  This stacking method simulates the measurement of flux-averaged colors, which effectively weights brighter sources higher than fainter sources relative to a logarithmic mean.  We use observed $V$ and $J$ colors, which correspond to nearly $u$ and $R$ at $z\sim0.8$.  Using a single circular, annular aperture of $0\farcs75$ inner radius and $1\farcs5$ outer radius, the observed $V_{AB}-J_{AB}$ colors of the stacked soft and hard samples are $0.81$ and $0.95$ magnitudes, respectively, suggesting that the soft sources are $0.14$ magnitudes bluer than the hard sources beyond $r \sim 6$ kpc.   Although this value is larger than the difference in $u-R$ colors measured via SED averaging, the similarity is reasonable considering that (1) the bands are in the observed frame and not the rest-frame and (2) a single aperture is being used for a variety of host morphologies.  

We have seen that the mean rest-frame $V-J$ color is also similar for the soft and hard sources (Figure \ref{fig:dust_plots}).  This implies that the mean rest-frame $u-R$ colors of the soft and hard subsamples are affected by dust extinction at similar levels.  Given both the similarity of the rest-frame $u-R$ colors and the similarity of the mean dust extinction in the soft and hard subsamples, it follows that the mean extinction-corrected rest-frame $u-R$ colors of the soft and hard sources would also be similar.

\section{DISCUSSION AND CONCLUSION}

\subsection{Comparison with Previous Work}

This work is the first to measure the UV/optical colors of a large sample of AGN host galaxies at $z \sim 1$ while quantitatively, reliably correcting for point source contamination in individual galaxies.  Ignoring or insufficiently correcting for point source contamination results in systematic error in comparing the colors of hard and soft X-ray sources (see Figure \ref{fig:UR_CMD_vectors}).

Direct comparisons of our results with those obtained for Seyfert galaxies in the local universe are problematic, as numerous AGN selection techniques are commonly used in the literature.  At low redshift, the effects of point source contamination on host galaxy color measurements are negligible because the galaxies are well-resolved.  Locally, \citet{sch09} find little host color difference between obscured and unobscured X-ray-selected AGN, in agreement with our results.  Although the sample is small, their selection techniques and use of X-ray-derived obscuration measurements are more analogous to those utilized in this study.  Given the similarity of our selection criteria, the agreement of our results suggests that the mean color offset between soft and hard sources has not evolved since $z \sim 1$.

\citet{ho08} find that unobscured broad-line AGN hosts at low-redshift are observed to possess large reservoirs of HI gas.  This result is interpreted as being evidence against a feedback scenario, in which unobscured AGNs are thought to be caught in the act of expelling gas \citep{ho08}.  No comparison is made between Type I and II sources.  Considering the use of different selection methodologies and measurements, there is no inconsistency between this study and our own.  For a sample of mid-infrared Type II QSOs at $0.3 < z < 0.8$, \citet{lac07} find that the host galaxies exhibit significant star formation rates ($3-90\;M_{\sun}$ yr$^{-1}$) and that disk inclination correlates with silicate features, implying that at least some of the reddening arises from the host galaxy.  However, it is not clear how these star formation rates would compare to similarly selected Type I QSOs at similar redshift.  Moreover, the AGN selection techniques, selected luminosity range, and redshift range are different from those in the present study, preventing direct comparison.

\citet{pag04} compares the $850\;\mu $m fluxes of luminous ($L_X \sim 10^{45}$ ergs s$^{-1}$), X-ray-selected AGN at $1 < z < 3$, finding that obscured sources have significantly more submillimeter emission than unobscured sources.  As a function of redshift, the mean flux difference between obscured and unobscured sources is consistent with zero at redshifts below $z \sim 1.5$ and rises significantly at higher redshifts.  Contrastingly, \citet{sha10} measures the far-infrared-derived star formation rates of X-ray selected intermediate-luminosity AGNs ($L_X \sim 10^{43}$ ergs s$^{-1}$) at $z\sim1$ in GOODS-North, finding no dependence of star formation rate on X-ray absorbing column density.   It appears that the mean star formation rate of obscured sources diverges from that of unobscured sources at either the highest X-ray luminosities ($L_X \sim 10^{45}$ ergs s$^{-1}$), at higher redshifts beyond $z \sim 2$, or in both of these regimes.  

CP10 measures the rest-frame U-V colors of intermediate-luminosity, X-ray-selected AGN in the Extended Groth Strip \citep{pie10b}.  They find that soft sources are systematically bluer than hard sources in both nuclear ($r < 0\farcs2$), extended ($0\farcs2 < r < 1\farcs0$), and integrated apertures.  The mean difference in extended color between soft and hard populations in our GOODS sample differs from the CP10 value by $2.5\sigma$ ($0.3$ mag in $u-R$).  Our color offsets are consistent to within $1\sigma$ ($0.1$ mag in $u-R$) when the CP10 sample is trimmed of all sources that they identify visually as having ``definite'' or ``possible'' point sources.  In addition, color offsets are similar when no correction for point source contamination is used in our sample ($0.3$ mag in $u-R$; see Figure \ref{fig:UR_CMD} and Table \ref{tab:tabulated_results}).  In CP10's Figure 11d it is clear that sources visually identified as possessing ``definite'' or ``possible'' point sources display anomalous color gradients (i.e., blue nuclear regions).  These lines of evidence point to some amount of point source contamination in CP10's extended colors.

\citet{car10} find that X-ray selected AGN hosts at $z \sim 1$ are frequently in dust-enshrouded galaxies.  Approximately one-half of our sample occupy a region in the rest-frame $U-V$ vs. $V-J$ diagram corresponding to $A_V > 1$, so our results are not inconsistent.  Although the authors do not explicitly compare the locations of obscured and unobscured sources in this plot, it is clear from their Figure 2 that the mean locations of these populations are similar, as in our sample.

\subsection{Implications for AGN Unification and Galaxy Evolution Scenarios}

In this paper, we have assembled an X-ray-selected AGN sample and measured the mean $u-R$ colors beyond $r\sim6$ kpc in the host galaxies.  We found that the mean rest-frame SEDs were similar for soft and hard sources both for objects of intermediate X-ray luminosities ($L_X > 10^{42}$ ergs s$^{-1}$) and objects of higher luminosities ($L_X > 10^{43.5}$ ergs s$^{-1}$, $\langle L_X\rangle \sim 10^{43.9}$ ergs s$^{-1}$).   We also found that the mean loci of the soft and hard populations in a rest-frame $U-V$ vs. $V-J$ were similar, implying that the mean properties of galactic dust extinction of the two populations were similar.   In Section \ref{sect:ERS_colors}, we used high spatial resolution WFC3/IR imaging to compute the mean rest-frame $NUV-R$ color gradients of soft and hard sources.  The CLEAN deconvolution algorithm was used to remove PSF contamination effects before computing colors.  It was seen that the colors of soft and hard sources were significantly different only in the nucleus ($r < 1.3$ kpc), and when sources with strong point sources in the observed $V$-band were removed from the soft sample, the mean colors of the soft and hard sources were similar with respect to the error bars at all radii.  The limiting spatial resolution of the WFC3 imaging prevents us from probing colors at radii smaller than $r \sim 1$ kpc.  We conclude that the host colors of soft and hard sources are statistically equivalent at all radii larger than $r \sim 1$ kpc.  These observations indicate that unobscured AGN are not redder than obscured AGN at $z \sim 1$ at intermediate X-ray luminosities.

In our sample, galaxy-wide dust extinction is weakly or not correlated with nuclear AGN obscuration as probed by X-ray hardness ratio.  Taken with the evidence that galaxy color is weakly or not correlated with X-ray hardness ratio, this implies that the conditions of star formation and dust extinction are uncorrelated with the conditions of neutral hydrogen obscuration on nuclear scales.  These observations favor AGN unification scenarios \citep{ant93, urr95}, in which AGN obscuration is determined by the orientation of a torus on parsec scales with respect to the observer.  This conclusion appears to be independent of X-ray luminosity in our sample for luminosities lower than $10^{44.5}$ ergs s$^{-1}.$  

Recently, several models of galaxy formation have been able to explain the bimodality of galaxy colors through AGN feedback processes \citep[e.g.,][]{dim05, lap06, hop06, men08}.  These models imply that unobscured AGN would be associated with a quenching of star formation, resulting in a reddening of the stellar population, even in the outer regions of the host (beyond $r \sim 6$ kpc) in the most energetic cases \citep{hop06, men08}.   The color of a single stellar population (SSP) reddens by $\sim0.7$ magnitudes in $u-R$ over a quasar lifetime following the ceasing of star formation (solar metallicity from age 10 Myr to age 100 Myr, \citet{mar05}), and thus the difference in color resulting from quenching in unobscured hosts should have been detectable in this study.

Our observations instead suggest that the majority of intermediate luminosity AGN at $z\sim1$ are not undergoing nor have recently experienced rapid, galaxy-wide quenching due to AGN-driven feedback processes.   A second possible interpretation is that if AGN-related blowout events have occurred in the population of intermediate luminosity AGNs at $z\sim1$, the radial extent of quenched stellar populations must be restricted to $\sim1$ kpc on average or the color signature of the quenched stellar population must have since been rewritten by fresh gas accretion and resulting star formation.

\section{ACKNOWLEDGEMENTS}	

This work has been supported in part by the NSF Science and Technology Center for Adaptive Optics, managed by the University of California (UC) at Santa Cruz under the cooperative agreement No. AST-9876783.

Some of the data presented in this paper were obtained from the Multimission Archive at the Space Telescope Science Institute (MAST). STScI is operated by the Association of Universities for Research in Astronomy, Inc., under NASA contract NAS5-26555. Support for MAST for non-HST data is provided by the NASA Office of Space Science via grant NAG5-7584 and by other grants and contracts.

Support for Program number HST-HF-51250.01-A was provided by NASA through a Hubble Fellowship grant from the Space Telescope Science Institute, which is
operated by the Association of Universities for Research in Astronomy, Incorporated, under NASA contract NAS5-26555.

S.M.A acknowledges fellowship support by the Allen family through UC Observatories/Lick Observatory.  This work is based in part on observations made with the European Southern Observatory telescopes obtained from the ESO/ST-ECF Science Archive Facility.  Observations have been carried out using the Very Large Telescope at the ESO Paranal Observatory under Program ID(s): LP168.A-0485.

{}

\end{document}

%% file: tab1.tex
\begin{table}

\begin{center}
\caption{Table of sample sizes, redshift ranges, mean redshift, mean X-ray luminosities, summarized color information, and selection criteria for all subsamples.}
\label{tab:tabulated_results}
\begin{tabular}[t]{llllllll}
\tableline\tableline
\multicolumn{1}{l}{Sample} &\multicolumn{1}{l}{\#}& \multicolumn{1}{l}{$z$ range} & \multicolumn{1}{l}{$\langle z\rangle$} &  \multicolumn{1}{l}{$\langle L_X\rangle$} & \multicolumn{1}{l}{$\langle (u-r)_{int}\rangle$} & \multicolumn{1}{l}{$\langle (u-r)_{outer}\rangle$} & \multicolumn{1}{l}{Selection}	\\
\tableline	\\[6pt]
all					&78		&0.5-1.5	&0.877	&43.11	&1.65$\pm$0.09	&1.69$\pm$0.10	&	$L_X > 42$	\\
all-hard				&42		&0.55-1.5	&0.869	&43.10	&1.81$\pm$0.10	&1.74$\pm$0.11	&	$L_X > 42$, $HR > -0.2$	\\
all-soft				&31		&0.5-1.3	&0.864	&43.12	&1.45$\pm$0.11	&1.65$\pm$0.12	&	$L_X > 42$, $HR < -0.2$	\\[6pt]

lum-all			&19		&0.5-1.5	&1.018	&43.92	&1.31$\pm$0.12	&1.65$\pm$0.14	&	$L_X > 43.5$					\\
lum-hard			&8		&0.5-1.5	&0.994	&43.85	&1.94$\pm$0.17	&1.76$\pm$0.18	&	$L_X > 43.5$, $HR > -0.2$			\\
lum-soft			&10		&0.5-1.3795&1.001	&44.01	&0.78$\pm$0.16	&1.61$\pm$0.17	&	$L_X > 43.5$, $HR < -0.2$			\\[6pt]

ERS-all			&24		&0.5-1.5	&0.788	&43.21	&1.61$\pm$0.11	&1.78$\pm$0.13	&	$L_X > 42$					\\
ERS-hard			&15		&0.5-1.5	&0.793	&43.02	&1.85$\pm$0.15	&1.74$\pm$0.16	&	$L_X > 42$, $HR > -0.2$			\\
ERS-soft			&9		&0.5-1.5	&0.781	&43.54	&1.20$\pm$0.16	&1.85$\pm$0.17	&	$L_X > 42$, $HR < -0.2$			\\[12pt]
\tableline\tableline


\tableline
\end{tabular}
\end{center}
\end{table}

%% file: tab2.tex

\begin{center}
\begin{longtable}{lllllllll}
\caption{Tabulated redshifts, X-ray luminosities, hardness ratios, and measured colors for individual sources.  Sources identified with ``LUO'' are GOODS-South sources as numbered in \citet{luo10}.  Sources identified with ``XID'' are GOODS-South sources as numbered in \citet{szo04} but not found in \citet{luo10}.  Sources identified with ``ABB'' are GOODS-North sources numbered in \citet{ale03} with spectroscopic redshifts from \citet{bar03}. } \label{tab:data} \\

\hline \hline \\[-2ex]
   \multicolumn{1}{c}{\textbf{XID}} &
   \multicolumn{1}{c}{\textbf{z}} &
      \multicolumn{1}{c}{\textbf{type}} &
         \multicolumn{1}{c}{\textbf{$L_X$}} &
            \multicolumn{1}{c}{\textbf{HR}} &
   \multicolumn{1}{c}{\textbf{$(u-r)_{outer}$}} &  
      \multicolumn{1}{c}{\textbf{$(U-B)_{outer}$}} &  
         \multicolumn{1}{c}{\textbf{$(u-r)_{int}$}} &  
      \multicolumn{1}{c}{\textbf{$(U-B)_{int}$}}  \\[0.5ex] \hline
   \\[-1.8ex]
\endfirsthead

\multicolumn{3}{c}{{\tablename} \thetable{} -- Continued} \\[0.5ex]
  \hline \hline \\[-2ex]
   \multicolumn{1}{c}{\textbf{XID}} &
   \multicolumn{1}{c}{\textbf{z}} &
      \multicolumn{1}{c}{\textbf{type}} &
         \multicolumn{1}{c}{\textbf{$L_X$}} &
            \multicolumn{1}{c}{\textbf{HR}} &
   \multicolumn{1}{c}{\textbf{$(u-r)_{outer}$}} &  
      \multicolumn{1}{c}{\textbf{$(U-B)_{outer}$}} &  
         \multicolumn{1}{c}{\textbf{$(u-r)_{int}$}} &  
      \multicolumn{1}{c}{\textbf{$(U-B)_{int}$}}   \\[0.5ex] \hline
  \\[-1.8ex]
\endhead

  \multicolumn{3}{l}{{Continued on Next Page\ldots}} \\
\endfoot

  \\[-1.8ex] \hline \hline
\endlastfoot
\hline
\rule[-1ex]{0pt}{3.5ex}  LUO 408 & 1.23 & 1 & 43.79 & -0.45 & 1.66 & 0.564 & 
1.70 & 0.564\\
\hline
\rule[-1ex]{0pt}{3.5ex}  LUO 370 & 1.02 & 1 & 43.48 & 0.12 & 1.87 & 1.04 & 1.71
 & 0.694\\
\hline
\rule[-1ex]{0pt}{3.5ex}  LUO 335 & 1.22 & 1 & 43.46 & -0.2 & 2.18 & 0.695 & 
0.341 & 0.0121\\
\hline
\rule[-1ex]{0pt}{3.5ex}  LUO 324 & 0.84 & 1 & 44.03 & -0.53 & 2.07 & 0.628 & 
0.683 & 0.0259\\
\hline
\rule[-1ex]{0pt}{3.5ex}  LUO 319 & 0.66 & 1 & 43.04 & -0.49 & 2.48 & 0.716 & 
1.39 & 0.435\\
\hline
\rule[-1ex]{0pt}{3.5ex}  LUO 316 & 0.67 & 1 & 43.36 & -0.4 & 1.93 & 0.723 & 1.99
 & 0.665\\
\hline
\rule[-1ex]{0pt}{3.5ex}  LUO 302 & 0.84 & 1 & 43.21 & -0.32 & 1.09 & 0.559 & 
1.46 & 0.529\\
\hline
\rule[-1ex]{0pt}{3.5ex}  LUO 288 & 1.03 & 0.5 & 43.44 & -0.37 & 1.48 & 0.521 & 
1.78 & 0.582\\
\hline
\rule[-1ex]{0pt}{3.5ex}  LUO 283 & 0.96 & 0.5 & 43 & -0.23 & 0.944 & 0.225 & 
1.28 & 0.341\\
\hline
\rule[-1ex]{0pt}{3.5ex}  LUO 273 & 0.74 & 1 & 43.52 & -0.56 & 1.64 & 0.679 & 
0.722 & 0.241\\
\hline
\rule[-1ex]{0pt}{3.5ex}  LUO 267 & 1.22 & 1 & 44.19 & -0.47 & 1.83 & 0.669 & 
0.459 & 0.0749\\
\hline
\rule[-1ex]{0pt}{3.5ex}  LUO 249 & 0.67 & 1 & 43.29 & 0.52 & 1.77 & 0.566 & 1.82
 & 0.669\\
\hline
\rule[-1ex]{0pt}{3.5ex}  LUO 247 & 0.73 & 1 & 44.48 & -0.54 & 1.83 & 0.669 & 
0.380 & 0.0914\\
\hline
\rule[-1ex]{0pt}{3.5ex}  LUO 246 & 0.73 & 1 & 43 & 0.06 & 2.00 & 0.669 & 2.01 & 
0.662\\
\hline
\rule[-1ex]{0pt}{3.5ex}  LUO 242 & 1.03 & 1 & 44.23 & -0.63 & 1.08 & 0.454 & 
0.386 & 0.135\\
\hline
\rule[-1ex]{0pt}{3.5ex}  LUO 224 & 0.73 & 1 & 43.23 & 0.44 & 2.07 & 0.771 & 1.67
 & 0.609\\
\hline
\rule[-1ex]{0pt}{3.5ex}  LUO 216 & 0.53 & 1 & 42.78 & -0.47 & 2.23 & 0.657 & 
2.38 & 0.704\\
\hline
\rule[-1ex]{0pt}{3.5ex}  LUO 168 & 1.1 & 1 & 44.21 & 0.61 & 1.65 & 0.755 & 1.91
 & 0.698\\
\hline
\rule[-1ex]{0pt}{3.5ex}  LUO 167 & 0.57 & 1 & 43.17 & -0.55 & 1.91 & 0.588 & 
1.74 & 0.546\\
\hline
\rule[-1ex]{0pt}{3.5ex}  LUO 145 & 0.67 & 1 & 42.92 & -0.45 & 2.24 & 0.794 & 
2.20 & 0.783\\
\hline
\rule[-1ex]{0pt}{3.5ex}  LUO 127 & 0.61 & 1 & 43.42 & 0.11 & 1.88 & 0.589 & 1.45
 & 0.470\\
\hline
\rule[-1ex]{0pt}{3.5ex}  LUO 49 & 1.04 & 1 & 43.73 & -0.44 & 2.22 & 0.789 & 
0.620 & 0.168\\
\hline
\rule[-1ex]{0pt}{3.5ex}  LUO 270 & 0.96 & 1 & 43.32 & -0.54 & 1.43 & 0.644 & 
1.45 & 0.458\\
\hline
\rule[-1ex]{0pt}{3.5ex}  LUO 311 & 1.31 & 1 & 42.64 & -1 & 0.316 & 0.0924 & 
0.668 & 0.168\\
\hline
\rule[-1ex]{0pt}{3.5ex}  LUO 118 & 1.03 & 1 & 42.8 & 0.1 & 1.89 & 0.715 & 1.74
 & 0.613\\
\hline
\rule[-1ex]{0pt}{3.5ex}  LUO 228 & 1.09 & 1 & 43.56 & 1 & 1.90 & 0.764 & 1.99 & 
0.752\\
\hline
\rule[-1ex]{0pt}{3.5ex}  LUO 189 & 0.6 & 1 & 43.3 & 1 & 1.86 & 0.643 & 2.28 & 
0.679\\
\hline
\rule[-1ex]{0pt}{3.5ex}  LUO 90 & 0.55 & 1 & 42.49 & 0.16 & 1.61 & 0.604 & 1.78
 & 0.583\\
\hline
\rule[-1ex]{0pt}{3.5ex}  LUO 381 & 0.66 & 1 & 42.32 & -0.17 & 2.07 & 0.719 & 
2.20 & 0.752\\
\hline
\rule[-1ex]{0pt}{3.5ex}  LUO 304 & 0.84 & 1 & 42.84 & 0.01 & 1.13 & 0.271 & 1.57
 & 0.333\\
\hline
\rule[-1ex]{0pt}{3.5ex}  LUO 415 & 1.14 & 1 & 42.42 & -1 & 1.92 & 0.712 & 2.11
 & 0.757\\
\hline
\rule[-1ex]{0pt}{3.5ex}  LUO 393 & 0.67 & 1 & 43.9 & 0.39 & 2.39 & 0.833 & 2.27
 & 0.774\\
\hline
\rule[-1ex]{0pt}{3.5ex}  LUO 109 & 0.93 & 0.9 & 42.83 & 0.13 & 1.35 & 0.531 & 
1.67 & 0.619\\
\hline
\rule[-1ex]{0pt}{3.5ex}  LUO 204 & 0.73 & 1 & 42.38 & 1 & 2.09 & 0.766 & 2.46 & 
0.796\\
\hline
\rule[-1ex]{0pt}{3.5ex}  LUO 310 & 0.73 & 1 & 43.22 & 1 & 2.09 & 0.748 & 2.05 & 
0.717\\
\hline
\rule[-1ex]{0pt}{3.5ex}  LUO 211 & 1.22 & 1 & 42.65 & -1 & 1.08 & 0.374 & 1.45
 & 0.462\\
\hline
\rule[-1ex]{0pt}{3.5ex}  LUO 254 & 0.74 & 1 & 42.19 & -1 & 1.99 & 0.654 & 2.29
 & 0.804\\
\hline
\rule[-1ex]{0pt}{3.5ex}  LUO 115 & 0.76 & 0.6 & 42.17 & -0.47 & 1.44 & 0.331 & 
2.03 & 0.451\\
\hline
\rule[-1ex]{0pt}{3.5ex}  LUO 386 & 1.17 & 1 & 43.43 & 0.52 & 1.11 & 0.462 & 1.71
 & 0.579\\
\hline
\rule[-1ex]{0pt}{3.5ex}  LUO 226 & 1.04 & 1 & 43.23 & 0.23 & 1.52 & 0.746 & 1.27
 & 0.554\\
\hline
\rule[-1ex]{0pt}{3.5ex}  LUO 260 & 1.32 & 1 & 43.45 & 0.6 & 1.08 & 0.289 & 0.834
 & 0.244\\
\hline
\rule[-1ex]{0pt}{3.5ex}  LUO 131 & 0.74 & 1 & 43.54 & 1 & 1.61 & 0.599 & 1.89 & 
0.631\\
\hline
\rule[-1ex]{0pt}{3.5ex}  LUO 72 & 0.72 & 1 & 43.39 & 1 & 1.78 & 0.377 & 1.75 & 
0.357\\
\hline
\rule[-1ex]{0pt}{3.5ex}  XID 511 & 0.77 & 1 & 42.01 & -0.24 & 0.868 & 0.379 & 
0.952 & 0.411\\
\hline
\rule[-1ex]{0pt}{3.5ex}  LUO 298 & 0.67 & 1 & 42 & -0.14 & 1.98 & 0.694 & 2.13
 & 0.743\\
\hline
\rule[-1ex]{0pt}{3.5ex}  XID 516 & 0.67 & 1 & 42.36 & -0.26 & 0.726 & 0.330 & 
1.02 & 0.371\\
\hline
\rule[-1ex]{0pt}{3.5ex}  LUO 235 & 1.03 & 1 & 42.62 & -0.03 & 0.930 & 0.452 & 
1.05 & 0.419\\
\hline
\rule[-1ex]{0pt}{3.5ex}  LUO 178 & 0.96 & 1 & 42.46 & 0.18 & 1.93 & 0.667 & 1.85
 & 0.642\\
\hline
\rule[-1ex]{0pt}{3.5ex}  LUO 117 & 0.68 & 1 & 42.49 & 0.34 & 2.14 & 0.748 & 2.17
 & 0.774\\
\hline
\rule[-1ex]{0pt}{3.5ex}  LUO 114 & 0.58 & 1 & 42.41 & -0.01 & 1.72 & 0.498 & 
2.00 & 0.632\\
\hline
\rule[-1ex]{0pt}{3.5ex}  XID 580 & 0.66 & 1 & 42.32 & -1 & 1.90 & 0.730 & 2.21
 & 0.726\\
\hline
\rule[-1ex]{0pt}{3.5ex}  LUO 129 & 1.33 & 1 & 43.72 & 1 & 1.61 & 0.472 & 1.95 & 
0.551\\
\hline
\rule[-1ex]{0pt}{3.5ex}  LUO 177 & 0.74 & 1 & 44.37 & 1 & 2.03 & 0.821 & 2.28 & 
0.839\\
\hline
\rule[-1ex]{0pt}{3.5ex}  LUO 195 & 0.67 & 1 & 43.3 & 1 & 2.01 & 0.694 & 2.28 & 
0.777\\
\hline
\rule[-1ex]{0pt}{3.5ex}  XID 611 & 0.98 & 1 & 43.41 & 1 & 1.38 & 0.646 & 2.00 & 
0.635\\
\hline
\rule[-1ex]{0pt}{3.5ex}  XID 612 & 0.74 & 1 & 43.3 & 1 & 1.49 & 0.527 & 1.68 & 
0.570\\
\hline
\rule[-1ex]{0pt}{3.5ex}  ABB 40 & 1.379 & 1 & 44.42 & -0.43 & 1.42 & 0.299 & 
1.24 & 0.228\\
\hline
\rule[-1ex]{0pt}{3.5ex}  ABB 55 & 0.64 & 1 & 42.04 & -0.1 & 1.67 & 0.567 & 1.71
 & 0.576\\
\hline
\rule[-1ex]{0pt}{3.5ex}  ABB 82 & 0.68 & 1 & 42.6 & 0.34 & 1.73 & 0.642 & 1.93
 & 0.723\\
\hline
\rule[-1ex]{0pt}{3.5ex}  ABB 90 & 1.14 & 1 & 43.42 & 0.5 & 1.57 & 0.575 & 1.72
 & 0.504\\
\hline
\rule[-1ex]{0pt}{3.5ex}  ABB 113 & 0.85 & 1 & 43.2 & -0.39 & 2.01 & 0.783 & 2.19
 & 0.670\\
\hline
\rule[-1ex]{0pt}{3.5ex}  ABB 115 & 0.68 & 1 & 43.49 & -0.34 & 1.71 & 0.663 & 
1.19 & 0.423\\
\hline
\rule[-1ex]{0pt}{3.5ex}  ABB 121 & 0.52 & 1 & 42.35 & 0.2 & 1.73 & 0.476 & 1.92
 & 0.524\\
\hline
\rule[-1ex]{0pt}{3.5ex}  ABB 142 & 0.75 & 1 & 42.61 & -0.07 & 2.21 & 0.814 & 
2.22 & 0.775\\
\hline
\rule[-1ex]{0pt}{3.5ex}  ABB 150 & 0.76 & 1 & 42.89 & 0.49 & 1.50 & 0.582 & 1.79
 & 0.667\\
\hline
\rule[-1ex]{0pt}{3.5ex}  ABB 157 & 1.26 & 1 & 43.99 & 0.38 & 1.15 & 0.341 & 1.31
 & 0.365\\
\hline
\rule[-1ex]{0pt}{3.5ex}  ABB 158 & 1.01 & 1 & 43.11 & 0.34 & 2.09 & 0.725 & 2.23
 & 0.737\\
\hline
\rule[-1ex]{0pt}{3.5ex}  ABB 177 & 1.02 & 1 & 42.66 & -0.32 & 1.19 & 0.432 & 
1.54 & 0.504\\
\hline
\rule[-1ex]{0pt}{3.5ex}  ABB 193 & 0.96 & 1 & 43.67 & -0.55 & 0.878 & 0.362 & 
0.742 & 0.279\\
\hline
\rule[-1ex]{0pt}{3.5ex}  ABB 201 & 1.02 & 1 & 42.8 & 0.64 & 1.91 & 0.786 & 1.73
 & 0.601\\
\hline
\rule[-1ex]{0pt}{3.5ex}  ABB 205 & 1.38 & 0.5 & 43.51 & -0.33 & 1.21 & 0.259 & 
1.55 & 0.314\\
\hline
\rule[-1ex]{0pt}{3.5ex}  ABB 212 & 0.94 & 1 & 42.26 & -0.37 & 2.19 & 1.08 & 1.80
 & 0.756\\
\hline
\rule[-1ex]{0pt}{3.5ex}  ABB 217 & 0.52 & 1 & 42.2 & 0.68 & 2.63 & 0.768 & 2.51
 & 0.721\\
\hline
\rule[-1ex]{0pt}{3.5ex}  ABB 222 & 0.86 & 1 & 42.93 & -0.22 & 1.90 & 0.654 & 
1.96 & 0.664\\
\hline
\rule[-1ex]{0pt}{3.5ex}  ABB 351 & 0.94 & 1 & 42.14 & -0.3 & 1.77 & 0.650 & 1.99
 & 0.680\\
\hline
\rule[-1ex]{0pt}{3.5ex}  ABB 352 & 0.94 & 1 & 42.55 & 0.53 & 1.49 & 0.547 & 1.61
 & 0.587\\
\hline
\rule[-1ex]{0pt}{3.5ex}  ABB 384 & 1.02 & 1 & 43.52 & 0.77 & 1.74 & 0.648 & 1.95
 & 0.675\\
\hline
\rule[-1ex]{0pt}{3.5ex}  ABB 451 & 0.84 & 1 & 44.04 & -0.42 & 1.44 & 0.568 & 
0.893 & 0.260\\

 \hline

\end{longtable}
\end{center}